\documentclass[12pt,nofootinbib]{article}

\usepackage[dvips]{graphicx}
\usepackage{amssymb}
\usepackage{amsmath}
\usepackage{epsfig}
\usepackage{amsfonts}
\usepackage{color}
\usepackage{mathtools}
\usepackage{enumerate}
\usepackage{comment}
\definecolor{White}{rgb}{1,1,1}
\definecolor{Black}{rgb}{0,0,0}
\usepackage{sidecap}
\usepackage{epsf}
\usepackage{graphicx,epsfig}
\usepackage{amsfonts}
\usepackage{amssymb}
\usepackage{tabularx,latexsym,alltt,graphicx,textcomp,hyperref,amsmath,amssymb,color}

\usepackage{soul}
\definecolor{dm}{cmyk}{.20, 0, .30, 0}
\sethlcolor{dm}

\makeatletter
\renewcommand\section{\@startsection {section}{1}{\z@}%
                                 {-3.5ex \@plus -1ex \@minus -.2ex}
                                   {2.3ex \@plus.2ex}%
                                   {\normalfont\large\bfseries}}
\renewcommand\subsection{\@startsection{subsection}{2}{\z@}%
                                   {-3.25ex\@plus -1ex \@minus -.2ex}%
                                     {1.5ex \@plus .2ex}%
                                     {\normalfont\bfseries}}
\renewcommand\subsubsection{\@startsection{subsubsection}{3}{\z@}%
                                   {-3.25ex\@plus -1ex \@minus -.2ex}%
                                     {1.5ex \@plus .2ex}%
                                     {\normalfont\itshape}}
\makeatother

\newcommand{\Letter}{
\setlength{\textwidth}{16.5cm}
   \setlength{\textheight}{22.8cm}
    \hoffset=-0.5in
\voffset=-2.1cm }

\Letter

\newcommand{\be}{\begin{equation}}
\newcommand{\ee}{\end{equation}}
\newcommand{\bea}{\begin{eqnarray}}
\newcommand{\eea}{\end{eqnarray}}
\newcommand{\barr}{\begin{array}}
\newcommand{\earr}{\end{array}}
\newcommand{\bl}{\left(}
\newcommand{\br}{\right)}

\newcommand{\bls}{\left[}
\newcommand{\brs}{\right]}
\def\te{\text}

\def\beq{\begin{equation}}
\def\eeq{\end{equation}}
\def\be{\begin{equation}}
\def\ee{\end{equation}}
\def\bea{\begin{eqnarray}}
\def\eea{\end{eqnarray}}

\newcommand{\eq}[1]{Eq.~(\ref{#1})}

\newcommand{\cit}[1]{Ref.~(\cit{#1})}

\begin{document}
\begin{titlepage}
\baselineskip=15.5pt
\pagestyle{plain}
\setcounter{page}{1}

\bigskip\
\begin{center}
{\Large \bf D-brane Bremsstrahlung}
\vskip 15pt
\end{center}
\vspace{0.5cm}
\begin{center}
{ Thomas C. Bachlechner and Liam McAllister}

\vspace{0.2cm}

\vskip 4pt
\textsl{Department of Physics, Cornell University,
Ithaca, NY USA 14853}\\
\vskip 4pt
\end{center}
\small  \noindent  \\[0.2cm]
\noindent
\vspace{1cm}
\hrule \ \\
\noindent {\bf Abstract} \\
We study the dynamics of ultrarelativistic D-branes. The dominant phenomenon is bremsstrahlung: mild acceleration induced by closed string interactions triggers extremely rapid energy loss through radiation of massless closed strings. After characterizing bremsstrahlung from a general $k$-dimensional extended object in a $D$-dimensional spacetime, we incorporate effects specific to D-branes, including  velocity-dependent forces and open string pair creation.  We then show that dissipation due to bremsstrahlung can substantially alter the dynamics in DBI inflation.

\bigskip
\hrule
\vfill \today
\end{titlepage}
\newpage

\tableofcontents

\begin{center}
\line(1,0){350}
\end{center}

\section{Introduction}

In the application of string theory to cosmological model-building,
moving D-branes have played a distinctive role.  For a probe D-brane in nonrelativistic motion in a fixed supergravity background, the Dirac-Born-Infeld and Chern-Simons actions govern
the evolution, and the dynamics is well-understood.
However, violent motions of branes, including relativistic collisions, are commonplace in cosmological models,
and relativistic D-brane scattering differs in important ways from the scattering of point particles or of strings.
In particular, massless bremsstrahlung from an extended object is qualitatively different from that from a point particle, and pair production of massive open strings stretched between colliding D-branes can substantially alter the dynamics.

In this paper we study ultrarelativistic D-brane scattering, incorporating closed string exchange, radiation of massless
closed strings, and pair production of massive open strings.
We find that supergravity interactions of widely-separated D-branes
induce modest
acceleration, which then triggers intense bremsstrahlung of massless closed strings, leading to rapid deceleration.
Open string pair production \cite{Bachas:1992bh,Bachas:1995kx,McAllister:2004gd} generally sets in at much smaller separations than bremsstrahlung, and can therefore be neglected early in the scattering process.

One interesting implication of our findings is that bremsstrahlung provides an important source of dissipation in the DBI scenario  \cite{Silverstein:2003hf,Alishahiha:2004eh}, and can alter the  evolution of the background and perturbations.  We find that for a range of reasonable parameter values, bremsstrahlung dramatically affects the trajectory.
However, there are also controllable parameter regimes in which bremsstrahlung can be neglected and the DBI model receives no corrections from our considerations.

The organization of this paper is as follows.
In \S\ref{sec:bremsstrahlung}  we determine the rate of energy loss from bremsstrahlung for a $k$-dimensional extended object undergoing an accelerated motion in a $D$-dimensional spacetime (for $D$ even).
Then, in \S\ref{sec:openandclosed} we compute the effect of bremsstrahlung on relativistic D-brane scattering, and demonstrate that a very small acceleration induced by velocity-dependent interactions triggers massive energy losses to closed string radiation.
In \S\ref{sec:applications} we
discuss implications for DBI inflation, and we conclude in \S\ref{sec:conclusions}.
In Appendix \ref{sec:onlyopenstrings} we characterize open string pair production, which can become important if the scattering D-branes reach a small separation while remaining relativistic.  In Appendix \ref{wkb} we present a few details  of the calculation of the critical frequency  for synchrotron radiation in an even-dimensional spacetime.

\section{Bremsstrahlung from Extended Objects}\label{sec:bremsstrahlung}

Two D-branes approaching each other at relativistic speeds experience a radial potential from closed string exchange
(see \S\ref{sec:openandclosed}).  As we shall see, the mild acceleration that results triggers
intense bremsstrahlung\footnote{We remark that the process termed `brane bremsstrahlung' in Ref.~\cite{Brax:2009hd}
corresponds to production of {\it{open}} string modes as a consequence of the masses of these modes being time-dependent.
In the present work  bremsstrahlung has the standard meaning of radiation emitted by an accelerated object.}
in the forward direction, substantially decelerating the branes.
In this section we will characterize this effect in general, determining the rate at which energy is lost to bremsstrahlung
when a $k$-dimensional extended object moving in a $D$-dimensional spacetime undergoes a specified acceleration.
In generic situations in which the forces in the directions parallel to and transverse to the direction of
motion are of the same order, radiation due to  transverse acceleration  is dominant, as  in synchrotron radiation in four dimensions.   However, in certain cases the force
can be nearly parallel to the velocity --- for example,  this arises in DBI inflation  with a purely radial
potential, see \S\ref{sec:applications} ---  and in this situation the radiation due to longitudinal acceleration can dominate.
In this work we will consider both purely transverse and purely longitudinal acceleration.

There are several important differences between the present problem and the familiar case of synchrotron radiation in four-dimensional electromagnetism:
scalar and gravitational radiation play an important role, the accelerated source is a $k$-dimensional\footnote{Radiation in four dimensions from particles corresponding to D-branes wrapping cycles of a toroidal compactification has been studied in Ref.~\cite{AbouZeid:1999fs}, while here we determine the radiation from spatially extended D-branes.} extended object, and the ambient spacetime is $D$-dimensional.  Even so, to provide intuition we will make extensive use of  comparisons to electromagnetism.

We begin in \S\ref{generalformalism} with the general formalism for spin-$s$ radiation by a point particle, determining the total power and spectral distribution.
In \S\ref{extendedradiation} we use this information to obtain the contribution of high-frequency  radiation to the total power radiated by a $k$-dimensional extended object, considering the special cases of transverse and parallel acceleration
in an even-dimensional\footnote{Odd-dimensional spacetimes present additional technical complications, essentially due to the fact that the Green's function has support in the interior of the past light cone,
and will not be treated here.  Although we assume Minkowski spacetime throughout this work, our results remain valid for more general spacetimes with curvature radii that are large compared to the typical wavelength of the radiation.}
Minkowski spacetime. It turns out that in the ultrarelativistic limit the radiation is phase-space suppressed, necessitating a careful treatment of the low frequency spectrum,  which we present in \S\ref{sec:classicalapprox}. Finally, in \S\ref{Hagedorn} we  consider the emission of massive strings by an accelerating D-brane.

Throughout our analysis, the primary large number in the system  will be the Lorentz factor $\gamma = {\rm cosh}\,\eta$,  where $\eta$  is the rapidity.
Correspondingly, a key task is to determine the scaling with $\gamma$ of the radiated power density $d P/dV_{k}$:
\begin{equation}
d P/dV_{k} \propto \gamma^{\Theta(D,k,s)}\,,
\end{equation}
where the exponent $\Theta$ may depend on $k$, $D$, and the spin $s$ of the field being radiated, and $V_{k}$ denotes the volume of the  extended object.

\subsection{General formalism for rank $s$ radiation from a point source}\label{generalformalism}

The first step is to determine the total power that a point source moving on a general trajectory in $D$ dimensions loses to radiation of integer spin $s$. Consider a field $A_{\mu_{1}...\mu_{s}}$ that is sourced  by a point of charge $e$
moving along  some given trajectory $r^{\mu}(\tau)$ with D-velocity $v^{\mu}=\partial_{\tau}r^{\mu}$, where $\tau$ is the source's proper time. Following Ref.~\cite{Mironov:2007nk,Kazinski:2002mp}, the  corresponding wave equation can be written
\be
\Box A_{\mu_{1}\dots \mu_{s}}(x)= J_{\mu}(x),~~~~~~ J_{\mu}(x')=e \int d\tau~ \delta^{D}[x'-x(\tau)]v_{\mu_{1}}\dots v_{\mu_{s}}\,.\label{waveeq}
\ee
The solution to \eq{waveeq} is given in terms of the $D$-dimensional retarded Green's function $D_{r}(x-r)$ as
\be
A_{\mu_{1}\dots \mu_{s}}(x)=e \int d\tau ~D_{\text{r}}(x-r) v_{\mu_{1}}\dots v_{\mu_{s}}\,.
\ee
Up to an overall $D$-dependent prefactor, which we will suppress throughout, the retarded Green's function in even dimensions is
\be
D_{\text{r}}(x-r)=\theta(x_{0}-r_{0})\delta^{(D/2-2)}\left((x-r)^{2}\right)\,.
\ee
where the exponent $(D/2-2)$  on the delta function  denotes a corresponding derivative with respect to the argument.
For the field $A_{\mu_{1}\dots \mu_{s}}$ one finds
\bea
A_{\mu_{1}\dots \mu_{s}}(x)&=&e \int d\tau ~\theta(x_{0}-r_{0}) {v_{\mu_{1}}\dots v_{\mu_{s}} \over{d\over d\tau}(x-r)^{2}}\frac{d}{d\tau}\delta^{(D/2-3)}\left((x-r)^{2}\right)\nonumber\\
&=&{e\over 2^{D/2-1}} \left({1\over R\cdot v} {d\over d\tau}\right)^{D/2-2}{v_{\mu_{1}}\dots v_{\mu_{s}} \over{R\cdot v}}\,,
\eea
where we performed $D/2-2$ partial integrations, and defined $R^{\mu}=(x-r)^{\mu}=(R,R {\bf n})$.
In our conventions, the $R$ in any dot product refers to the $D$-vector $R^{\mu}$, while $R=|{\bf x}-{\bf r}|$ otherwise.  Note that any contribution from derivatives of the Heaviside step function involves a factor $\delta(x_{0}-r_{0})$. Combined with the second delta function, there is no contribution unless $R=x_{0}-r_{0}=|{\bf x}-{\bf r}|=0$, so that we can ignore this contribution away from the source.

In a similar way one can show that to leading order in $1/R$,
\be
\partial_{\mu}A_{\mu_{1}\dots \mu_{s}}(x)=-{e\over 2^{D/2-1}} R_{\mu} \left({1\over R\cdot v} {d\over d\tau}\right)^{D/2-1}{v_{\mu_{1}}\dots v_{\mu_{s}} \over{R\cdot v}}\,.\label{eq:radfield}
\ee
To examine the stress energy tensor, we work in the Lorenz gauge, defined by $\partial^{\mu_{s}}A_{\mu_{1}\dots \mu_{s}}=0$.  At leading order in $1/R$ and neglecting derivatives of $R$ (see also Ref.~\cite{Mironov:2007nk}),
\be
\partial^{\mu_{s}}A_{\mu_{1}\dots \mu_{s}}=e \left({1\over R\cdot v} {d\over d\tau} \right)^{D/2-1}v_{\mu_{1}}\dots v_{\mu_{s-1}}\ne 0 ~~\te{for }s\ge2\,.\label{eq:gaugefixing}
\ee
This indicates that for $s\ge 2$ the forces  responsible for driving the particle on the assumed trajectory $r^{\mu}(\tau)$ will have significant backreaction on the total power. This effect is well-known in general relativity;  our result agrees with Refs.~\cite{Peters:1972ec,Price:1973ns},
where the amount of gravitational radiation due to stresses on the source is found to be comparable to the radiation from the source  itself.

Neglecting for the moment the  radiation due  to forces acting on the source, we find
the stress-energy tensor
\be
T_{\mu \nu}^{s}=e^{2}R_{\mu}R_{\nu}\left[ \left({1\over R\cdot v} {d\over d\tau}\right)^{D/2-1}{v_{\mu_{1}}\dots v_{\mu_{s}} \over{R\cdot v}}\right]^{2}\, .\label{eq:emt}
\ee
The power radiated per unit solid angle is given by
\be
{d P_{D,s}\over d\Omega_{D-2}}=\lim_{R\rightarrow \infty} R^{D-2}\left[ T_{0i}n^{i}\right] {R\cdot v\over R\gamma}\, ,\label{eq:angulardistr}
\ee
which reduces to the well-known Larmor formula for $s=1$.

To find the spectrum, note that the total energy radiated, $I^{0}_{D,s}$, obeys\footnote{See also Refs.~\cite{Cardoso:2002pa, Cardoso:2007uy}.}
\bea
{d I^{0}_{D,s}\over d\Omega_{D-2}}&=&\int dt~\lim_{R\rightarrow \infty} R^{D-2}\left[ T_{0i}n^{i}\right]_{\te{ret}}\nonumber\\
&=&\lim_{R\rightarrow \infty} R^{D} \int d\omega ~|e A_{D}(\omega)|^{2}\, ,\label{specdistr}
\eea
where we have defined
\be
A_{D}(\omega)=\int dt~e^{i\omega t}\left[ \left({1\over R\cdot v} {d\over d\tau}\right)^{D/2-1}{v_{\mu_{1}}\dots v_{\mu_{s}} \over{R\cdot v}}\right]_{\text{ret}}\,.
\ee
The spectrum of radiation is then given by
\be
{d^{2} I^{0}_{D,s}\over d\omega d\Omega_{D-2}}=\lim_{R\rightarrow \infty} R^{D}|e A_{D}(\omega)|^{2}\,.\label{eq:pointspectrum}
\ee
To bring this into a useful form, we recall that the derivatives are evaluated at the retarded time $t=t'+R(t')\approx t'+x-{\bf n}\cdot {\bf r}(t')$. Changing variables of integration, we have\footnote{Useful relations for this evaluation are $R\cdot v=R\gamma (1-{\bf n}\cdot {\boldsymbol \beta}(t'))$ and $dt/dt'=1-{\bf n}\cdot {\boldsymbol \beta}(t')$.} \cite{Mironov:2007nk}
\be
A_{D}(\omega)=\int dt'~e^{i\omega (t'-{\bf n}\cdot {\bf r}(t'))} {R\cdot v\over \gamma R} \left[ \left({1\over R\cdot v} {d\over d\tau}\right)^{D/2-1}{v_{\mu_{1}}\dots v_{\mu_{s}} \over{R\cdot v}}\right]\,.\label{eq:freqamp}
\ee
To obtain the energy spectrum in any even dimension we perform $D/2-2$ partial integrations in \eq{eq:freqamp} and use $d/d\tau=\gamma d/dt'$ to arrive at\footnote{One subtlety is that \eq{eq:ampintegrated} only applies when  the integration is taken over all time, so that the boundary terms in the partial integration vanish. This is a legitimate assumption for transverse acceleration  of a source that remains on a circular orbit, but is not  appropriate for a purely longitudinal acceleration: see \S\ref{linearcase}.}
\be
A_{D}(\omega)=\omega^{(D-4)/2}R^{1-D/2} \int dt'~e^{i\omega (t'-{\bf n}\cdot {\bf r}(t'))} {d\over d\tau}{v_{\mu_{1}}\dots v_{\mu_{s}} \over{R\cdot v}}\,.\label{eq:ampintegrated}
\ee
The remaining integral is, up to angular dependence, just the Fourier-transformed amplitude in four dimensions. Combining \eq{eq:ampintegrated} and \eq{eq:pointspectrum}, we find the energy spectrum
\be
{d^{2} I^{0}_{D,s}\over d\omega d\Omega_{D-2}}=\lim_{R\rightarrow \infty} \omega^{D-4}\left[ e A_{4}(\omega)\right]^{2}\,,\label{eq:evenspectrum}
\ee
which relates the energy spectrum radiated in four dimensions to the energy spectrum in any even dimension, up to angular dependence and numerical prefactors. Note in particular that the critical frequency above which the spectrum decreases exponentially is independent of the  spacetime dimension $D$.\footnote{We have checked this result by numerically evaluating the power spectrum for scalar and vector radiation in diverse spacetime dimensions.
Further analytical evidence  comes from a WKB approach to the radiated power spectrum, following Ref.~\cite{Breuer}, as detailed in Appendix \ref{wkb}.}

\subsection{Radiation by a $k$-dimensional source in $D$ dimensions} \label{extendedradiation}
Having understood radiation from a point particle in $D$ dimensions, we now characterize the power density
$dP^{k}_{D,s}/dV_{k}$ of spin $s$ radiation from a $k$-dimensional extended object in $D$ dimensions.

To obtain the energy spectrum radiated from an extended object we need to integrate the Fourier-transformed field, \eq{eq:radfield}, over the $k$ spatial dimensions parallel to the object:
\be
\partial_{\mu}A_{\mu_{1}\dots \mu_{s}}(\omega)=\int dV_{k}~\rho \sin[\omega(t+\phi({\bf r_{k}}))] R_{\mu}A_{D}(\omega)\, ,
\ee
where $\phi$ is a phase,  ${\bf r}_{k}$ parameterizes the position along the extended object, $\rho$ is the charge density of the source, and $A_{D}$ is the radiation amplitude in frequency space given in \eq{eq:freqamp}.
The resulting radiated energy spectrum is
\bea
{d^{2}I^{k}_{D,s}\over d V_{k}d{\omega}}=\lim_{R\rightarrow \infty} R^{D-k} \int dV_{D-2-k}\left\langle \left|\partial_{\mu}A_{\mu_{1}\dots \mu_{s}}(\omega)\right|^{2}\right\rangle,\label{eq:intensity}
\eea
where the brackets indicate a time average.
Now let ${\bf r}_{0}$ parameterize the position of an arbitrary fixed point on the extended object,
so that we can write ${\bf{r}}={\bf r}_{0}(t')+{\bf r}_{k}$.
The time average of the squared amplitude then takes the form
\begin{eqnarray}
\begin{split}
&\left\langle \left|\partial_{\mu}A_{\mu_{1}\dots \mu_{s}}(\omega)\right|^{2}\right\rangle\\
&=\lim_{R\rightarrow \infty} \rho^{2}\left\langle\left| \int dt'dV_{k} \sin[\omega(t'+\phi({\bf r_{k}}))]~e^{i\omega (t'-{\bf n}\cdot {\bf r}({\bf r}_{k},t'))}{R\cdot v\over \gamma R} \left[ \left({1\over R\cdot v} {d\over d\tau}\right)^{D/2-1}{v_{\mu_{1}}\dots v_{\mu_{s}} \over{R\cdot v}}\right] \right|^{2}\right\rangle \, .
\end{split}
\end{eqnarray}
Considering radiation that is peaked in the forward direction, we can model the angular dependence of the amplitude by the dimensionless function $f(\Omega)\sim e^{-|\theta|/\theta_{c}}$, where $\theta_{c}$ is the angle beyond which emission is negligible, and we assume isotropic radiation along the parallel directions. Denoting the solid angle within which emission occurs as $\Omega_{c}$, we have
\be
\left\langle \left|\partial_{\mu}A_{\mu_{1}\dots \mu_{s}}(\omega)\right|^{2}\right\rangle\approx[\rho A_{D}(\omega)]^{2}_{\Omega=\Omega_{c}}\left\langle\left| \int dV_{k}~ \sin[\omega(t+\phi({\bf r_{k}}))]f(\Omega)\right|^{2}\right\rangle\,,\label{eq:timeav}
\ee  where we have kept the leading terms in $\omega {\bf r}_{k}$. The time average in \eq{eq:timeav} can be evaluated in spherical coordinates, using   $\phi({\bf r_{k}}) \approx r/(2R)$ and $\theta\approx r/R$:
\bea
\left\langle\left| \int dV_{k}~ \sin(\omega(t+\phi({\bf r_{k}})))f(\Omega)\right|^{2}\right\rangle&=&\left\langle \left|\int_{V_{k}}d\Omega_{k-1}dr r^{k-1}{\sin\bls\omega \bl t+\sqrt{R^{2}+r^{2}}\br\brs}e^{-\theta(r)/\theta_{c}}\right|^{2}\right\rangle\nonumber\\
&\approx& {\lambda^{k} R^{k} \over 2 }\, .\label{timeaverage}
\eea
Combining the time average in \eq{timeaverage} with \eq{eq:intensity} and approximating the angular distribution as uniform for $\theta<\theta_{c}$, we find that for $\omega \gg \frac{2\pi}{\theta_{\te{c}}^2 R}$,
\be
{d^{2}I^{k}_{D,s}\over d V_{k}d{\omega}}\approx {\lambda^{k}\over 2 \theta_c^{k}}{\rho^{2}\over e^{2}} {d I^{0}_{D,s}\over d\omega }\, ,\label{extfull}
\ee
with $\lambda=2\pi/\omega$.
The result \eq{extfull} is physically sensible: two points on the brane that are separated by a distance $d$ radiate coherently at wavelength $\lambda$ if and only if $d \ll \lambda$.
Thus, the full power radiated at wavelength $\lambda$ is the sum of the power radiated by patches of volume $\lambda^{k}$,
leading to the factor of $\lambda^{k}$ in \eq{extfull}.\footnote{We thank H.~Tye for helpful explanations of this point.}  The dependence on the critical angle arises by observing that interference effects do not depend on $\theta_{c}$, so that a factor $\theta_{c}^{-k}$ from the point particle angular distribution remains in the final result.

\subsubsection{Circular trajectories}

In the previous section we obtained general expressions for radiation from sources on arbitrary trajectories.
We now use these results to characterize the  spectrum  of scalar, vector and gravitational synchrotron radiation
emitted by an ultrarelativistic source on a circular trajectory of radius $r_0$ in $D$ dimensions.
Note that there are additional subtleties for the case of gravitational radiation,
as the gauge-fixing condition \eq{eq:gaugefixing} is not satisfied (see also Ref.~\cite{Price:1973ns}).

The spectrum of spin $s$ radiation from an accelerated point particle of charge $e$ in four-dimensional spacetime is well known and can be summarized as  \cite{Breuer,Peters:1972ec,Price:1973ns}:
\be
{dP_{4,s}^{0}\over  d\omega}=e^{2} \omega_{0} \gamma \bl{\omega\over\omega_{\te{c}}}\br^{1-2s/3}e^{-2\omega/(3\omega_{\te{c}})}\,,\label{ansatz}
\ee
where $\omega_{0}$ is the angular frequency of the circular trajectory, and the critical frequency is given by $\omega_{c}=\gamma^{3}\omega_{0}$. The critical angle $\theta_{\te{c}}$ within which most of the radiation is emitted is \bea\label{equ:critangle}
\theta_{\te{c}}\sim\begin{cases}\left(\frac{\omega_0}{\omega}\right)^{1/3}~~&\text{for~} \omega_0\ll\omega\ll\omega_c \, ;\\\gamma^{-1}~~&\text{for~} \omega\sim\omega_c\,.\end{cases}\,
\eea
Combining \eq{ansatz} with \eq{eq:evenspectrum}, the power spectrum of a point particle in an even-dimensional spacetime is
\be
{dP_{D,s}^{0}\over  d\omega}=e^{2} \omega_{0} \theta_{c}^{D-4} \omega^{D-4} \gamma \bl{\omega\over\omega_{\te{c}}}\br^{1-2s/3}e^{-2\omega/(3\omega_{\te{c}})}\, .\label{eq:pointspectrum2}
\ee
Integrating over all frequencies, we have
\be
P_{D,s}^{0}\sim e^{2}(\gamma^{2}\omega_{0})^{D-2}\, ,
\ee
which is just the $D$-dimensional Larmor formula. Equipped with the power spectrum \eq{eq:pointspectrum2} for a point particle we can use \eq{extfull} to obtain the power spectrum of a $k$-dimensional extended object:
\be
{d^{2}P_{D,s}^{k}\over dV_{k}d\omega}\sim {(2\pi)^{k}\rho^{2}\over 2} \omega_{0} (\theta_{c} \omega)^{D-4-k} \gamma \bl{\omega\over\omega_{\te{c}}}\br^{1-2s/3}e^{-2\omega/(3\omega_{\te{c}})}\, .
\ee
Finally, to obtain a lower bound on the power density radiated by an extended object we will only take the high frequency\footnote{The power density from low-frequency radiation is not necessarily insignificant, but is more subtle, as we will explain.} contribution into account:
\be
{dP_{D,s}^{k}\over dV_{k}}=\int_{0}^{\infty}d{\omega}~ {d^{2}P_{D,s}^{k}\over dV_{k} d\omega}\gtrsim\int_{\omega_{0}}^{\infty}d{\omega}~ {d^{2}P_{D,s}^{k}\over dV_{k} d\omega}\, .\label{extendedrad}
\ee
Using Eqs.~(\ref{ansatz}), (\ref{equ:critangle}), and (\ref{extfull}) in \eq{extendedrad} we find that the power density lost to radiation is (omitting numerical coefficients)
\bea
{d P^{k}_{D,s}\over dV_{k}}&\gtrsim&
\rho^2 \gamma  \omega_{0} \int_{\omega_0}^\infty d\omega~ (\theta_{c} \omega)^{D-4-k}   \left(\frac{\omega}{\omega_c}\right)^{1-2s/3}e^{-2\omega/(3\omega_c)}\nonumber \\
&\approx & \rho^{2}(\gamma^{2}\omega_{0})^{D-2-k}\, .
 \label{closedproduc}
\eea
This is a key result for our analysis: the power of $\gamma$ that appears in \eq{closedproduc} controls the importance of bremsstrahlung in relativistic D-brane dynamics. Note that the power density emitted from a $k$-dimensional object in $D$-dimensional space scales exactly like the power from a point particle in $D-k$ dimensions, as expected from the symmetries of the problem.

\subsubsection{Linear acceleration} \label{linearcase}

Having understood the comparatively straightforward case of circular motion, we now turn to computing the total power radiated by an extended object undergoing linear acceleration, as well as the cutoff frequency $\omega_{c}$ of this radiation and the critical angle $\theta_{c}$ within which the radiation is confined.

We consider a particle with constant acceleration as measured in the comoving frame: this could be a particle in a constant electric field, as considered in Ref.~\cite{Reville:2010yy}. The potential is given by
\be
V(x)=V_{1}x.
\ee
The  resulting hyperbolic trajectory is given in terms of the proper time $\tau$ as
\be
x={m\over V_{1}} \left(\cosh\left(\tau { V_{1}\over m}\right) -1\right),~~~t={m\over V_{1}} \sinh\left(\tau { V_{1}\over m}\right)\, .
\ee
In particular, the acceleration is $a=V_{1}/(m\gamma^{3})$.

To  understand some of the subtle aspects of linear acceleration emission, consider a non-relativistic particle that enters a linear potential at $t=0$   with velocity $v(0)$ and leaves it with velocity $v(T)$ at time $t=T$. Following Ref.~\cite{Schwinger2}, the spectral distribution of the energy radiated by this particle is given by
\be
{dI^{0}_{4,1}\over d\omega}={2e^{2}\over 3\pi} \frac{1}{\omega^{2}} |{\bf \ddot{v}}(\omega)|^{2}\, .
\ee
This leads to the spectral distribution
\be
{dI^{0}_{4,1}\over d\omega}={8e^{2}\over 3\pi} {V_{1}^{2}\over m^{2}}{1\over \omega^{2}} \sin^{2}\left({\omega T\over 2}\right)\, .
\ee
Only frequencies of order $\omega_{c}\sim 1/T$ contribute to the spectrum. In particular, in the limit of constant acceleration, $T\rightarrow \infty$, one has $\omega_{c}\rightarrow 0$, which corresponds to a constant electric field, rather than to radiation.
Even so, the total energy radiated per unit time is
\be
{I^{0}_{4,1}\over T}={2e^{2} \over 3} {V_{1}^{2}\over m^{2}}={2e^{2} \over 3} a^{2}\, ,
\ee
which is just the Larmor formula. As Schwinger put it, {\it ``Facetiously, we may say that a uniformly accelerated charge radiates because it is not uniformly accelerated''} \cite{Schwinger2}. This summarizes an important (and long-studied) subtlety of linearly accelerated systems: as the acceleration and rapidity change constantly, one cannot assign an instantaneously radiated power --- the spectrum crucially depends on the whole history of the particle's trajectory. In the following, we will carefully take this into account.

For simplicity we will  give explicit results for the power spectrum and angular distribution of {\it{scalar}}  radiation from linearly-accelerated point particles.  The generalization to vector radiation is  straightforward, but as  noted above,  the extension to gravitational radiation is nontrivial: one cannot consistently neglect radiation from the stresses that cause the acceleration.

The power radiated per unit solid angle is given by \eq{eq:angulardistr} with the energy momentum tensor \eq{eq:emt}.  Combining these two results for linear acceleration, we find that in four dimensions\footnote{One  readily finds analogous  results in any even dimension.}
\be
{d P^{0}_{4,0}\over d\Omega_{2}}=\frac{e^2 \cos^{2}(\theta_{1}) \cos^{2}(\theta_{2}) {\dot{\beta}^{2}}}{\gamma^{2}\left(1-\cos(\theta_{1}) \cos(\theta_{2}) \beta\right)^5}\,.\label{eq:power4d}
\ee
Using \eq{eq:power4d} we can solve for the critical
angle $\theta_{c}$ within which most of the radiation is confined in the ultrarelativistic limit: we find
$\theta_{c}\approx \sqrt{2} \sqrt{1-\beta}\approx 1/\gamma$. It is easy to check that the total power  agrees with the Larmor formula:
\be
P^{0}_{4,0}={d P^{0}_{4,0}\over d\Omega_{2}}\bigg|_{\Omega=\Omega_{c}}\theta_{c}^{2}\propto e^{2}\gamma^{8}\dot{\beta}^{2} \gamma^{-2}=e^{2}\gamma_{\te{max}}^{6}\dot{\beta}^{2}\, .
\ee
Note that the critical angle  can depend on the frequency.
Using  intuition from the case of transverse acceleration, we expect that the critical angle increases significantly at low frequencies. Furthermore, the radiation of scalars is peaked in the forward direction for any even spacetime dimension. This is different from vector radiation, which vanishes in the strictly forward direction but is still constrained to lie within a critical angle $\theta_{c}\sim 1/\gamma$.

\begin{figure}
  \centering
  \includegraphics[width=10cm]{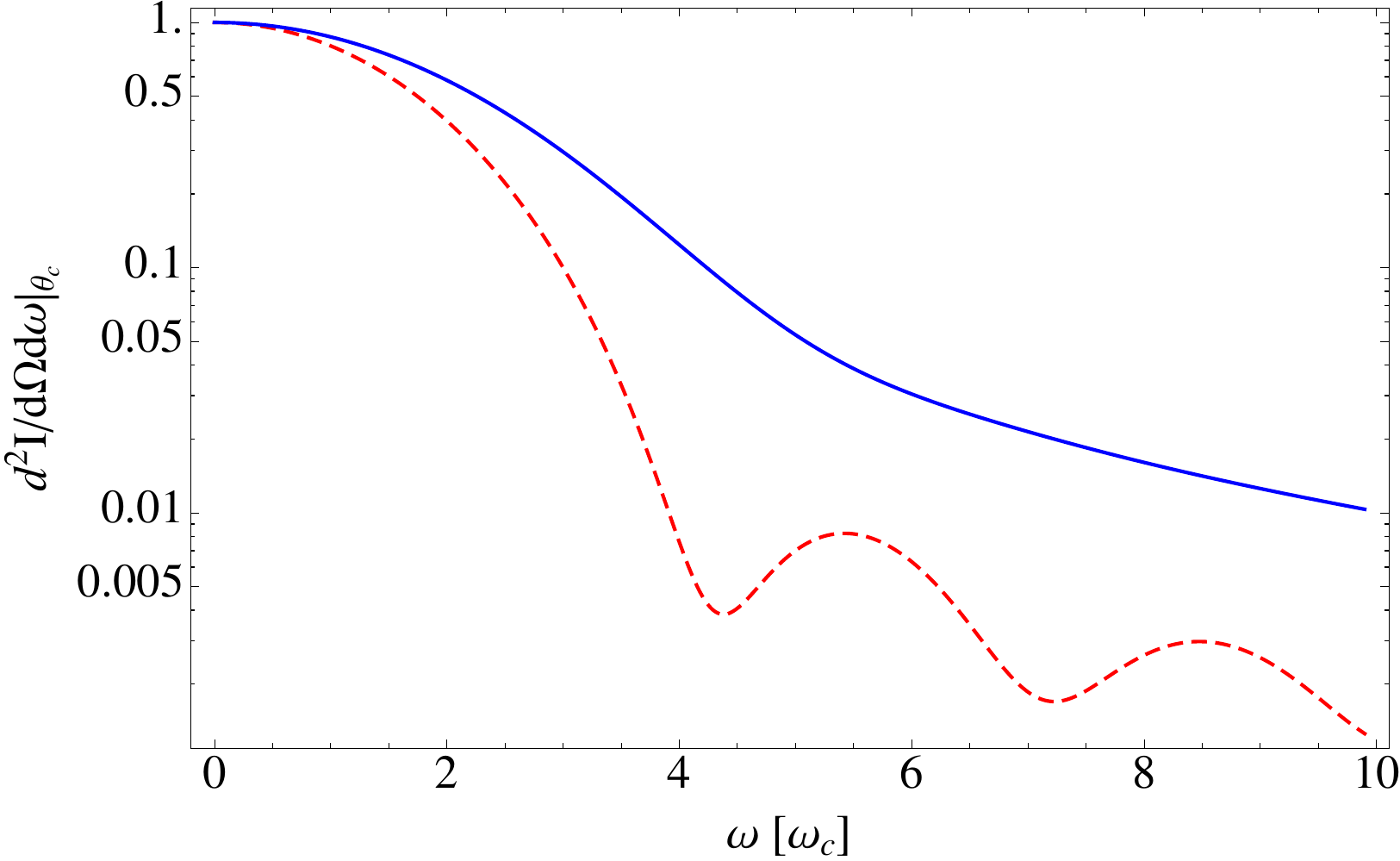}
  \caption{\small Radiated power spectrum from a point particle in four dimensions, \eq{eq:spect}. The solid blue line represents vector radiation, while the dashed red line represents scalar radiation.}\label{linearspectrum}
\end{figure}

To obtain the power spectrum we need to evaluate
\be
{d I^{0}_{4,0}\over d\omega d\Omega_{2}}=\lim_{R\rightarrow \infty} R^{4} |e A(\omega)|^{2}\, ,
\ee
where
\be
A_{4,0}(\omega)=\int_{t'=0}^{T} dt'~e^{i\omega (t'-{\bf n}\cdot {\bf r}(t'))} {R\cdot v\over \gamma R} \left[ \left({1\over R\cdot v} {d\over d\tau}\right)^{4/2-1}{1 \over{R\cdot v}}\right]\, .
\ee
Note that unlike the transverse acceleration case, we need to introduce a time cutoff $T$ at which we evaluate the radiation. This is because the spectrum constantly changes while the particle is accelerated. We therefore consider a particle  that accelerates linearly for $0<t<T$, and evaluate the spectrum at time $T$. We find
\be
{d I^{0}_{4,0}\over d\omega d\Omega_{2}}=e^{2}\left|\int_{0}^{T}dt'~ \frac{  \cos(\theta_{1}) \cos(\theta_{2}) }{\gamma \left(\cos(\theta_{1}) \cos(\theta_{2}) \beta-1\right)^2} e^{i \omega  (t'-\cos(\theta_{1}) \cos(\theta_{2}) x(t'))}\right|^{2}\, .\label{eq:spect}
\ee
To illustrate the spectrum of vector and scalar radiation from a point particle,  one can  evaluate \eq{eq:spect} numerically for linear acceleration in four dimensions. The resulting spectrum is shown in Figure~\ref{linearspectrum}.

A direct analytic evaluation of the integral in \eq{eq:spect} is difficult. Let us therefore examine the Fourier transform in the limit $\gamma\gg 1$ at $\theta_{1}=\theta_{2}=0$ and change variables to integration over $\gamma$. This gives
\bea
{d I^{0}_{4,0}\over d\omega d\Omega_{2}}&\approx&16 e^{2}\left| - e^{-\frac{i \omega m}{2 V_{1}}}+ e^{-\frac{i \omega m }{2V_{1}\gamma_{\te{max}}}}\gamma_{\te{max}}+\frac{ i \omega m \Gamma\left[0,\frac{i \omega m}{2 V_{1}}\right]}{2V_{1}}-\frac{ i \omega m \Gamma\left[0,\frac{i \omega m}{2 V_{1}\gamma_{\te{max}}}\right]}{2V_{1}}\right|^{2}\, .
\eea
Evaluated at $\omega=0$ we have ${d I^0_{4,0}\over d\omega d\Omega_{2}}\big|_{\omega=0}=16e^{2} \gamma_{\te{max}}^{2}$. On the other hand, expanding the power spectrum at high frequencies gives
\bea
{d I^{0}_{4,0}\over d\omega d\Omega_{2}}&\approx&{64e^{2} V_{1}^{2}\gamma_{\te{max}}^{4}\over m^{2} \omega^{2}}\, .
\eea
Now solving
\be
{d I^{0}_{4,0}\over d\omega d\Omega_{2}}\Big|_{\omega_{c}}\sim {1\over 2}{d I^0_{4,0}\over d\omega d\Omega_{2}}\Big|_{\omega=0}
\ee
gives the critical frequency
\be
\omega_{c}\sim {4 V_{1}\over m}\gamma_{\te{max}}= 4 a\gamma_{\te{max}}^{4}\, ,
\ee
where we inserted the acceleration from above. To perform a consistency check we can compute the total radiated power,
\be
{P^{0}_{4,0}}\approx \theta_{c}^{2} \omega_{c} {d\gamma_{\te{max}}\over dt}{d\over d\gamma_{\te{max}}}  {d I^{0}_{4,0}\over d\omega d\Omega_{2}}\Big|_{\omega_{c}} \sim {1\over \gamma_{\te{max}}^{2}} {V_{1}\over m}\gamma_{\te{max}} {V_{1}\over m} e^{2 } \gamma_{\te{max}}\sim e^{2}\gamma_{\te{max}}^{6} a^{2}\,,
\ee and recognize that this agrees precisely with the Larmor formula.

Finally, we are in a position to obtain a lower bound on the radiation emitted from extended objects. As in the previous section, we will only take into account high frequency radiation. However, due to the fact that the whole trajectory contributes to the radiation at a time $T$ and we cannot take $T\rightarrow \infty$, the result for the spectrum \eq{eq:evenspectrum} of a point source in higher dimension has limited applicability, and we do not have a simple expression for the critical angle $\theta_{c}(\omega)$ at small $\omega$. Due to these limitations we only consider frequencies $\omega\gtrsim \omega_{c} $, as this contribution does not depend  parametrically on the precise form of the spectrum at low frequencies.   We find
\bea
{dP_{D,s}^{k}\over dV_{k}}&=&\int_{0}^{\infty}d{\omega}~ {d^{2}P_{D,s}^{k}\over dV_{k} d\omega}\gtrsim\int_{\omega_{c}}^{\infty}d{\omega}~ {d^{2}P_{D,s}^{k}\over dV_{k} d\omega} \equiv \Biggl({dP_{D,s}^{k}\over dV_{k}}\Biggr)^{\te{high}}\nonumber\\ \label{extendedradlin}
&\sim& {\rho^{2}} (\gamma_{\te{max}}^{3} a)^{D-2-k}\, ,  \label{simpleextendedradlin}
\eea
where we  have used the facts that ${d P^{0}_{D,s}/ d\omega }\propto e^{-2\omega/(3\omega_{c})}$, $\theta_{c}(\omega_{c})=1/\gamma$ and $\omega_{c}=a\gamma_{\te{max}}^{4}$. We have checked numerically that these results hold for other, nonlinear potentials.

\subsection{Validity of the classical approximation}\label{sec:classicalapprox}

Until now we have considered classical radiation from a source on a fixed background trajectory.
For a point particle, the approximation of classical radiation breaks down when the energy carried by the emitted radiation
$\omega_{\te{c}}$ becomes comparable to the total energy\footnote{For an extended object the relevant energy scale depends on frequency, and is given by $E_{\te{max}}=\gamma \mu \lambda^{k}$, where $\mu$ is the tension of the object in its rest frame.}
$E_{\te{max}}= \gamma m$ of the source \cite{Reville:2010yy,Harko:2002eh}, so that the source recoils substantially upon emitting one
high-frequency quantum.
A useful parameter to characterize the importance of  quantum effects is $\zeta \equiv \omega_{c}/E_{\te{max}}$.
When $\zeta\ll 1$,  the classical  approximation  is applicable without correction, for the  entire range of frequencies in which the radiation is substantial.
On the other hand, for $\zeta\gg 1$ there is substantial radiation at frequencies  where quantum effects  are important,
and a proper treatment requires quantizing the fields  and computing the appropriate matrix element for radiation emission,
including the phase space factors. Such a calculation is beyond the scope of the present work.  Instead,
we will evaluate the classical amplitude at low energy, where it applies, and cut off the radiation spectrum below the
energy scale $E_{\te{max}}= \gamma m$ where quantum effects become significant.
This leads to a conservative lower bound on the radiation emitted.

\begin{figure}
  \centering
  \includegraphics[width=10cm]{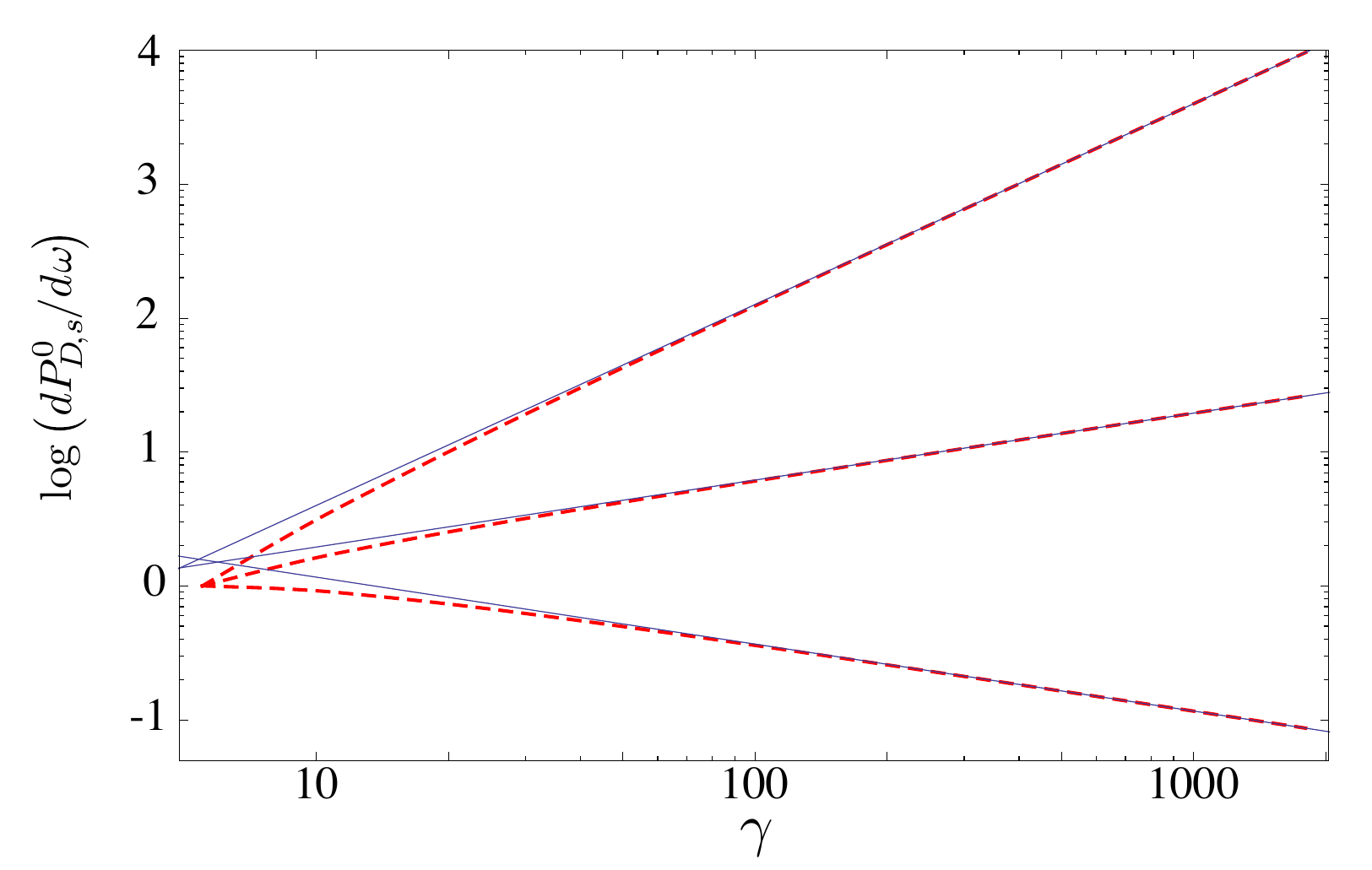}
  \caption{\small Power spectrum at low frequencies as a function of $\gamma$. The dashed red lines indicate the numerical evaluation of  \eq{powercut}, while the solid blue lines represent the estimate \eq{powerparticle}, up to a constant factor, for $D=4,6,8$. }\label{scalingcheck}
\end{figure}

Let us therefore place a lower bound on the total power emitted by an extended object undergoing linear acceleration
in the regime $\zeta\gg 1$, where the classical approximation of \S\ref{extendedradiation}
breaks down.\footnote{The case of transverse acceleration is conceptually similar and omitted for brevity.}
We cut off the energy spectrum at the maximum energy $E_{\te{max}}$, which  in general depends on  the frequency.
Using \eq{eq:evenspectrum},  one finds  that the
energy spectrum of a point particle at time $T$  is approximately
\be
{d^{2} I^{0}_{D,s}\over d\omega d\Omega_{D-2}}\sim\lim_{R\rightarrow \infty} \omega^{D-4}\left[ e A_{4}(\omega,T)\right]^{2}\theta\left[E_{\te{max}}(\omega)-\omega\right].
\ee
The total power radiated is then given by
\be
P^{0}_{D,s}=\partial_{T}\int_{\omega=0}^{E_{\te{max}}}d\omega  \int d\Omega_{D-2} ~\lim_{R\rightarrow \infty} \omega^{D-4}\left[ e A_{4}(\omega,T)\right]^{2}.\label{powercut}
\ee
Because the spectrum is approximately flat at low frequencies, and $E_{\te{max}}\ll \omega_{c}$ in the regime  of interest,
we can approximate the integral in \eq{powercut} as
\bea
P^{0}_{D,s}&=&E_{\te{max}} \theta_{c}^{D-2}{d\over dT}\left[\lim_{R\rightarrow \infty} \omega^{d-4}\left[ e A_{4}(\omega,T)\right]^{2}\right]_{\omega=E_{\te{max}}}\nonumber\\
&\sim& E_{\te{max}}e^{2}a^{D-3} \gamma^{3(D-2)-4},\label{powerparticle}
\eea
where we used  the fact that the emission cone has  opening angle $\theta_{c}$, which is evaluated at energy $E_{\te{max}}$.
To confirm the scaling with $\gamma$ we  have evaluated the integral numerically: the  numerical result is shown along with \eq{powerparticle} in Figure \ref{scalingcheck}.

Combining \eq{powerparticle} and \eq{extfull}, we find that the power density emitted by an extended object  takes the form
\begin{equation}
{dP^{k}_{D,s}\over dV_{k}}\sim \int d\Omega_{D-2}\int_{\omega=0}^{E_{\te{max}}} d\omega~ {1\over \omega^{k} \theta_{c}^{k}}{d^{2} P^{0}_{D,s}\over d\omega d\Omega_{D-2}}\,.
\end{equation}
To obtain a  very conservative\footnote{The estimate  is parametrically correct for all $k$,  but is conservative only  for $k>1$.}
parametric estimate of the power density, we may  perform  a small slice of the integral, in a region
$\omega_0 \ll \omega \ll E_{\te{max}}$, which gives
\bea
{dP^{k}_{D,s}\over dV_{k}} &\sim& E_{\te{max}} \theta_{c}^{D-2-k}{d\over dT}\left[\lim_{R\rightarrow \infty} \omega^{D-4-k}\left[ \rho A_{4}(\omega,T)\right]^{2}\right]_{\omega=E_{\te{max}}}\nonumber\\
&\sim& E_{\te{max}}\theta_{c}^{-k}E_{\te{max}}^{-k}{dP^{0}_{D,s}\over d\omega}\bigg|_{\omega=E_{\te{max}}}. \label{lowerbound}
\eea

We conclude that in the ultrarelativistic\footnote{Note that $\zeta \propto \gamma^2$.} limit, the  power density emitted by a linearly-accelerated object  with $k$  spatial dimensions has the parametric scaling
\bea\label{powerdens}
{dP^{k}_{D,s}\over dV_{k}}&\gtrsim& E_{\te{max}} {1\over E_{\te{max}}^{k} \theta_{c}^{k}(E_{\te{max}})} \rho^{2} a^{D-3} \gamma^{3(D-2)-4}=\left({E_{\te{max}}\over \omega_{c}}\right)^{1-k} \Biggl({dP_{D,s}^{k}\over dV_{k}}\Biggr)^{\te{high}}\bigg|_{\omega_{c}}\nonumber\\ &\approx&\left({\gamma\mu\over \omega_{c}^{k+1}}\right)^{1-k\over k+1} \Biggl({dP_{D,s}^{k}\over dV_{k}}\Biggr)^{\te{high}}\bigg|_{\omega_{c}}\,,
\eea
where the high-frequency contribution  was defined in \eq{simpleextendedradlin}, and we have  used $\theta_{c}(E_{\te{max}})\sim 1/\gamma$. For $k>1$ the  power in low-frequency radiation
is parametrically larger than the power in high-frequency radiation, as a result of coherence effects.
Note, however, that  this conclusion is valid only if the timescale over  which the trajectory of the radiating object changes significantly is much longer than the wavelength of  the radiation. Thus, the above estimate breaks down for very long wavelengths.

\subsection{Emission of massive strings}\label{Hagedorn}

Our results so far have described the massless radiation from a general extended object undergoing acceleration during ultrarelativistic motion.  However, quanta of massive fields of mass $m$ can be emitted when the spectrum of massless radiation reaches frequencies $\omega \gtrsim m$.  When the accelerating object is a D-brane,\footnote{For simplicity we state the results here for an accelerating D0-brane, but the generalization is trivial.} emission of excited closed strings could become significant if $\omega_c$ exceeds a critical value of order $1/\sqrt{\alpha^{\prime}}$, as we now show.
(Note that since $\omega_c = \gamma^3 \omega_0$, cf.~Appendix \ref{wkb}, $\omega_c \sim 1/\sqrt{\alpha^{\prime}}$ corresponds to $\omega_0 \ll 1/\sqrt{\alpha^{\prime}}$, and thus $a \ll 1/\sqrt{\alpha^{\prime}}$, so that the accelerations in question are far below the string scale.)
To obtain the power spectrum of a scalar field with mass $m$ we can use the results for a massless field and substitute $\omega\rightarrow \sqrt{\omega^{2}+m^2}$.
The power radiated in a field of mass $m^{2}=4n/\alpha^{\prime}$ therefore obeys\footnote{In this section we assume $\omega_{c}\ll E_{\te{max}}$, so that quantum effects do not play a role.}
\be
P > \int_{\omega_{\te{c}}}^{\infty}d\omega\left.{d P\over d\omega}\right|_{\omega\to\sqrt{\omega^{2}+4n/\alpha^{\prime}}}\gtrsim\int_{\sqrt{\omega_{\te{c}}^{2}+4n/\alpha^{\prime}}}^{\infty}d\omega\, {d P\over d\omega}\,.
\ee
Summing over the string spectrum, we obtain
\be
P > \sum_{n=0}^{\infty}d_{n}\int_{\sqrt{\omega_{c}^{2}+4n/\alpha^{\prime}}}^{\infty}d\omega~ {d P\over d\omega}\,,
\ee
where the level degeneracy $d_{n}$ for closed superstrings is (see Ref.~\cite{Sundborg:1984uk})
\be \label{dn}
d_{n} \sim { n^{-11/2}e^{4\pi\sqrt{2n}}}~~\te{for $n\gg 1$}\,.
\ee
Using
\be
\left.{d P\over d\omega}\right|_{\omega=\omega_c} \propto  \exp\bl{-\sqrt{1+{4n\over\alpha^{\prime}\omega^{2}_{\te{c}}}}}\br\,,\label{power1}
\ee and comparing to \eq{dn}, we learn that the emitted power diverges when $\omega_c > \omega_{\te{H}}$, where
$\omega_{\te{H}}$ is given by
\be
\omega_{\te{H}}={1\over 2\sqrt{2} \pi }\cdot\frac{1}{\sqrt{\alpha'}}\,.
\ee
We recognize that $\omega_{\te{H}}$ is precisely equal to the Hagedorn temperature $T_{\te{H}}$  for closed superstrings. In this estimate we assumed that the classical radiation power spectrum is valid, i.e.~$\omega_{\te{H}}\ll E_{\te{max}}$, which corresponds to $g_{s}\ll2^{3(k+1)/2}\pi^{1+k} \gamma$ and does not pose a strong constraint.

We conclude that a D-brane subject to an acceleration\footnote{The acceleration  $a_{\te{H}}$ given in \eq{aH} is parametrically smaller than the critical acceleration for fundamental strings obtained in Ref.~\cite{Parentani:1989gq}.}
\begin{equation} \label{aH}
a_{\te{H}}= \frac{1}{2\sqrt{2}\pi \gamma^{3}} =  \frac{T_{\te{H}}}{\gamma^{3}}
\end{equation} sources a divergent power density of bremsstrahlung of {\it{massive}} closed strings.
More physically, we expect that dissipation from bremsstrahlung will make it impossible for a D-brane to reach the Hagedorn acceleration $a_{\te{H}}$.
On the other hand, for $\omega_{\te{c}}\ll \omega_{\te{H}}$ the number of massive strings emitted is negligible, and the leading contribution to the radiated power comes from massless strings.

\section{D-brane Deceleration from Bremsstrahlung}\label{sec:openandclosed}
In the preceding section, we characterized bremsstrahlung from an accelerated extended object compelled by an  unspecified outside force to follow a {\it{fixed}} trajectory: that is, we neglected radiation reaction.
We will now specialize to  the case of a D-brane pair interacting via velocity-dependent forces, and
determine the radiation reaction force that results from the emission of bremsstrahlung.
We will find that radiation reaction dramatically alters the dynamics of relativistic D-branes.

For simplicity, we will work in the regime where the classical emission spectrum is valid, i.e. $\omega_{\te{c}} \ll E_{\te{max}}$, as discussed in \S\ref{sec:classicalapprox}.

\subsection{Interactions between moving D-branes}\label{sec:paircreation}

The first step is to characterize the interactions of D-branes in relativistic relative motion.  Consider a scattering process in which two D$k$-branes approach each other with impact parameter $b$ and initial relative velocity $v$.
As a first step we will study a simplified model in which the relative velocity $v$ remains constant, i.e.\ any energy loss due to bremsstrahlung or other interactions is not incorporated.  Of course, as our interest is in how losses affect a D-brane scattering event, we will ultimately discard this approximation.

The system of moving D-branes at constant relative velocity can be mapped to a pair of branes at rest that are misaligned by an imaginary angle $\phi$, which
is related to $v$ by $i\phi=\text{tanh}^{-1}(v)\equiv \eta$.  Evaluating the resulting annulus diagram is then a textbook problem \cite{polchinskitwo}.
T-dualizing as needed, one finds the following interaction amplitude for D$k$-branes that move relative to each other:
\be
{\cal A}=V_{\te{D}k} \int_0^{\infty} {dt \over t} (8 \pi^2 t)^{-k/2} \exp \left(-{{t b^2} \over {2 \pi }}\right) {\theta_{11}(\eta t/2 \pi, it)^4 \over {\theta_{11}( \eta t/\pi, it)\eta_{\te{D}}(it)^{9}}} \label{ampA}\,,
\ee
where
$V_{\te{D}k}$ is the volume of the D$k$-brane, and we have set
$\alpha^{\prime}=1$.  We have used the same definitions of the theta functions as in Ref.~\cite{McAllister:2004gd},  but we have denoted the Dedekind $\eta$  function by $\eta_{\te{D}}$ to avoid confusion with the rapidity $\eta$.

The real and imaginary parts of \eq{ampA} correspond to massless\footnote{Massive closed string exchange is Yukawa suppressed, and can be neglected at distances large compared to the string scale.} closed string exchange and open string pair production, respectively.  Closed string exchange becomes significant at much larger separations than those relevant for open string pair creation, and we defer a detailed treatment of open strings to Appendix \ref{sec:onlyopenstrings}.

To compute $\te{Re}({\cal A})$, we write ${\cal A} = \int_0^{\infty} dz\,A(z)$, and then consider a suitable closed contour in the complex $z$ plane.
We take the contour formed by the segments $\gamma_{1}$, $\gamma_{\mathbb{R}_{+}}$, and  $\gamma_{\te{Res}_{i}}$, where the ray $\gamma_{1}$ is parameterized by $R e^{i \epsilon}$, with $R\in(\infty, 0)$ and $\epsilon\ll 1$;
$\gamma_{\mathbb{R}_{+}}=\{x\in\mathbb{R}_{+}\backslash \{ n\pi/\eta,~n\in\mathbb{N}\}\}$; and $\gamma_{\te{Res}_{i}}$ is a small semicircle
above the $i$th singularity on the real axis.
Taken together with an arc at infinity, these segments parameterize a closed contour around a region that contains no poles, implying that
\be
0 = \te{Re}\bls\bl\int_{\gamma_{1}}+\int_{\gamma_{\mathbb{R}_{+}}} \br dz~A(z)-{2\pi i\over 2} \sum_{j}\te{Res}_{j\in \mathbb{N}}\bls A\bl {j \pi\over\eta} \br\brs\brs\,.
\ee
The residues of the poles along the real axis are purely real, so that
\be
\te{Re}\bls\int_{\gamma_{\mathbb{R}_{+}}} dz~A(z)\brs=-\te{Re}\bls\int_{\gamma_{1}} dz~A(z)\brs=\int_{R=0}^{\infty} dR~A(R e^{i \epsilon})\,,
\ee
where the imaginary part of the last integral vanishes because $\lim_{y\rightarrow 0}\te{Im}[A(x+i y)]=0$ for $x\ne n \pi/\eta$, $n\in \mathbb{N}$. It remains to evaluate the integral over $\gamma_{1}$, which is given by
\bea
\te{Re}({\cal A})&=&\int d\tau~ V(\tau)\nonumber\\
&=&\int d\tau\,dR\,V_{\te{D}k} {e^{-i \epsilon/2}(8 \pi^2 \varrho)^{-k/2} \over \sqrt{2\pi^{2}R} }  \exp \left(-{{\varrho r^{2}} \over {2 \pi }}\right) {\theta_{11}(\eta \varrho/2 \pi, i \varrho)^4 \over {\theta_{11}( \eta \varrho/\pi, i \varrho)\eta_{\te{D}}(it)^{9}}}\,, \label{vinttheta}
\eea
where  $\varrho = R e^{i \epsilon}$, and we have introduced dependence on a new variable, $\tau$, with $r^{2}=\tau^{2}\tanh^{2}(\eta)+b^2$.

Following Ref.~\cite{polchinskitwo}, $V(\tau)$ can be interpreted as the interaction potential in the frame in which one of the D-branes is at rest.
The integrand in \eq{vinttheta} decreases exponentially with $R$, and the characteristic scale of this decrease is given by $R_{\te{c}}\approx 2\pi/r^{2}$.
Moreover, the first pole along the real axis appears at $R_{\te{sing}}=\pi/\eta$.  Thus, for $\eta\ll r^{2}/2$ -- which we shall assume throughout our analysis -- the integral is well-approximated by integrating along the real axis up to some
$R_{\star}$ with $R_{\te{c}}\ll R_{\star}\ll R_{\te{sing}}$.

Using the asymptotic expression
\be
\theta_{11}(\eta t/\pi,i t)\approx\begin{cases} -\frac{2  \eta}{\sqrt{t}}\exp\bl {-\frac{\pi }{4 t}-\frac{t \eta^2}{\pi }}\br~~&\te{for $t\ll1$, $\eta\ll1$;}\\ -\frac{1}{\sqrt{t}}\exp\bl{-\frac{(\pi -2 t \eta)^2}{4 \pi  t}}\br~~&\te{for $t\ll1$, $\eta\gg1$\,,}\end{cases}
\ee the potential $V$ can be written as
\be
V(r)\approx -{2^{2-2k} \pi ^{5/2-3 k/2}} V_{\te{D}k}\Gamma \left[\frac{7-k}{2}\right] {\tanh(\eta) \eta^3 \over r^{7-k}} \label{lowvpot}
\ee
for $\eta\ll1$,
and
\be
V(r)\approx-2^{3-2 k}  \pi ^{5/2-3k/2}V_{\te{D}k}\Gamma\bls {7-k\over 2}\brs{\tanh(\eta)e^\eta\over r^{ 7-k}}\,\label{relpot}
\ee
for $1 \ll \eta \ll r^{2}$.  The nonrelativistic result \eq{lowvpot}
matches the result given in Ref.~\cite{polchinskitwo}, where the integrand is first expanded in small $\eta$ and then integrated.
However, the method used above also allows us to obtain the potential in the highly relativistic case, given in \eq{relpot}.

\subsection{D-brane deceleration}

Equipped with an understanding of the  forces between  relativistic D-branes, we can now  determine the
acceleration $a$ of the scattering branes, and then use \eq{closedproduc} to compute the power that is radiated into closed strings.

We work in the `fixed target' frame in which one D-brane is at rest, and we orient our coordinates so that the moving brane's initial velocity is parallel to $\hat{\bf x}$, with impact parameter $b \hat{\bf y}$.  As in the preceding sections, $\eta$ denotes the rapidity,  $ \eta \equiv \text{tanh}^{-1}(v)$, associated to the relative velocity of the two D-branes. The corresponding Lorentz factor is $\gamma = {\rm{cosh}}\,\eta \approx \frac{1}{2}e^{\eta}$.  We consider D$k$-branes with $0<k<7$, and set $D=10$.
Let $\dot{\rho}_{\text{closed}} = d P_{10}^k/dV_{k}$ denote the power density lost into radiation of massless closed strings, let $\rho_{\te{D}k} = \gamma T_{\te{D}k}$
denote the energy density of the moving D-brane, and let $V$ denote the interaction potential given in \eq{relpot}.

For small $v_{y}$, the acceleration in the $\hat{\bf x}$ direction can be approximated by $a_{x}\approx 4\dot{\eta} e^{-2\eta}\,.$  Using energy conservation one finds
\bea
\dot{\eta} \approx-\,{\dot{\rho}_{\text{closed}}+\dot{V}\over\rho_{\te{D}k}}\,.
\eea
The acceleration in the  $\hat{\bf y}$ direction is then fixed by angular momentum conservation, as we now explain.  We use a time variable $\tau$, with $\tau=0$ at the point of closest approach, and define $\tau_{\te{brem}}$ as the time at which the energy lost to bremsstrahlung becomes comparable to the initial kinetic energy.
A convenient parameterization is
\begin{equation}
\tau_{\te{brem}} \propto {\rm exp}\bigl(\Delta(k)\,\eta_0\bigr)\,,\label{deftb}
\end{equation} where $\Delta(k)$ is a constant that depends only on  $k$,
and we retain only factors exponential in the initial rapidity $\eta_0$.
Manifest in \eq{deftb} is the approximation that $\tau_{\te{brem}}$ is exponentially large, which we confirm below is self-consistent: see \eq{closedstop1}.
Specifically, we make the consistent approximation that
$|\tau_{\te{brem}}| \gg 1$, and that correspondingly $V \ll \rho_{\te{D}k}$.
With these preliminaries, one readily finds that
\be \label{ayis}
a_{y}\sim-{b\over \tau}{\dot{V}\over\rho_{\te{D}k}}\, .
\ee
Comparing $a_{x}$ and $a_{y}$, we find that the former is suppressed by a factor of $e^{-2\eta}$, so that unless $b\lesssim\tau e^{-2\eta_{0}}$ (corresponding to an exponentially small regime of parameters that we will omit in the remainder), the acceleration in the transverse direction dominates.\footnote{Note that the radiation  due to transverse acceleration is dominant only for impact parameters $b\lesssim\tau e^{-\eta_{0}}$, because of the higher power of $\gamma$ in the Larmor formula for linear acceleration.}

The analysis leading to \eq{ayis} neglected the angular momentum stored in any open string pairs that are produced, which is an excellent approximation:
we show in Appendix \ref{sec:onlyopenstrings} that the density of open strings is exponentially small.  On the other hand, bremsstrahlung does carry away angular momentum, but this effect cancels in \eq{ayis}.

Combining \eq{ayis} with the closed string interaction potential \eq{relpot} and using \eq{closedproduc}, we finally obtain the rate at which energy density is lost due to bremsstrahlung:
\be
\dot{\rho}_{\text{closed}}\approx
\sigma(k)G_{10}T_{\te{D}k}^{k-6} b^{8-k} \tau ^{(9-k) (k-8)}\,{\rm{exp}}\Bigl(2(8-k)\eta\Bigr)\,, \label{rhoclosed}
\ee
where $\sigma(k) \approx 2^{(k-8)(2k-3)}  \pi ^{(k-8) (3 k-5)/2} \Gamma[ {9-k\over 2}]^{8-k}$, and we remind the reader that we have set $\alpha'=1$.
Equation (\ref{rhoclosed}) is one of our main results. The power density lost to radiation scales with the Lorentz factor $\gamma$ in the fixed target frame as $\dot{\rho}_{\text{closed}} \propto \gamma^{2(8-k)}$,
from which it follows that {\it{bremsstrahlung provides an extremely strong dissipative effect in ultrarelativistic D-brane scattering.}}
Note that this result only provides a lower limit on the power density radiated:
only the high frequency part of the spectrum, with $\omega > \omega_c$, was used in obtaining \eq{rhoclosed}.  Moreover, we have not included acceleration induced by radiation reaction, which will in general induce additional bremsstrahlung.

Knowing the power radiated in massless closed strings,\footnote{As the acceleration is much smaller than the Hagedorn acceleration found in \S\ref{Hagedorn}, emission of massive closed strings is negligible.} we can determine the timescale for deceleration from bremsstrahlung,
i.e.\ the time $\tau_{\te{brem}}$ at which the rapidity is reduced by one: $\eta(\tau_{\te{brem}}) \equiv \eta_0-1 $.
Integrating \eq{rhoclosed} and setting $\eta \to \eta_0$ one finds\footnote{\eq{closedstop}, \eq{eta}, and \eq{eta1} hold for $\tau\gg b$ and receive corrections for $\tau\sim b$.}
\be
\tau_{\te{brem}}\approx-\left({2\sigma(k)\over (8-k)(9-k)-1} G_{10}T_{\te{D}k}^{k-7} b^{8-k} e^{(15-2k)(\eta_{0}-1)} \right)^{\frac{1}{(8-k) (9-k)-1}}\,.\label{closedstop}
\ee
We learn in particular that
\begin{equation}
\Delta(k) = \frac{15-2k}{(8-k)(9-k)-1}\,,\label{closedstop1}
\end{equation} so that $\tau_{\te{brem}}$ {\it{is an exponentially long time in the ultrarelativistic limit}}, as anticipated in \eq{deftb}.
Correspondingly, bremsstrahlung first becomes significant when the D-branes are very widely separated.

\begin{figure}
  \centering
  \includegraphics[width=10cm]{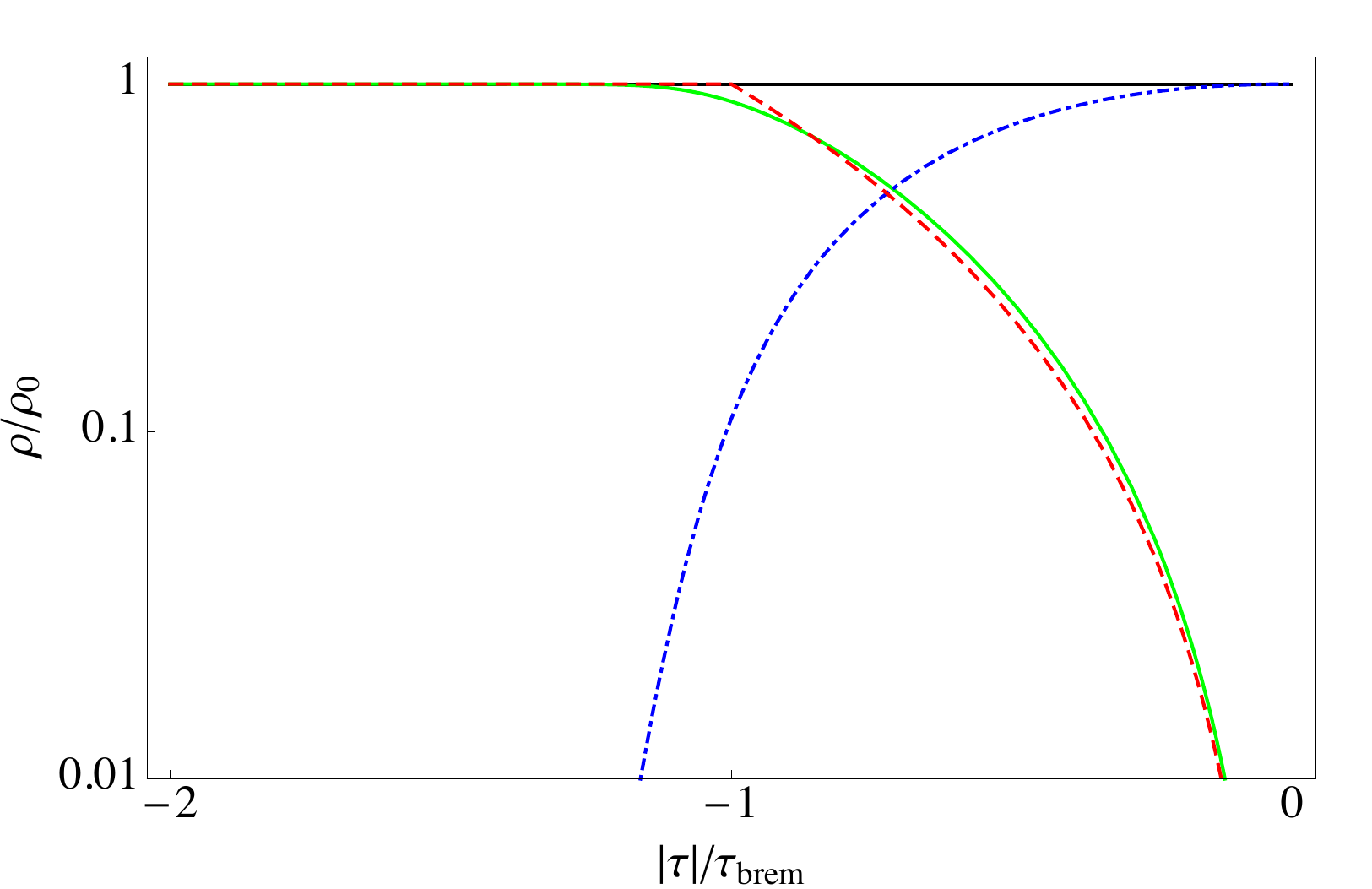}
\caption{\small Normalized energy, versus time measured in units of $\tau_{\te{brem}}$, for $\eta_{0}=26$ and $b=10\,l_{\te{s}}$.  The solid green curve shows the kinetic energy of the branes, while the  dot-dashed blue curve indicates the energy lost to closed string radiation. The dashed red line marks the analytical estimate, obtained from \eq{eta}, of the kinetic energy of the branes.}\label{short1}
\end{figure}

\begin{figure}
  \centering
  \includegraphics[width=10cm]{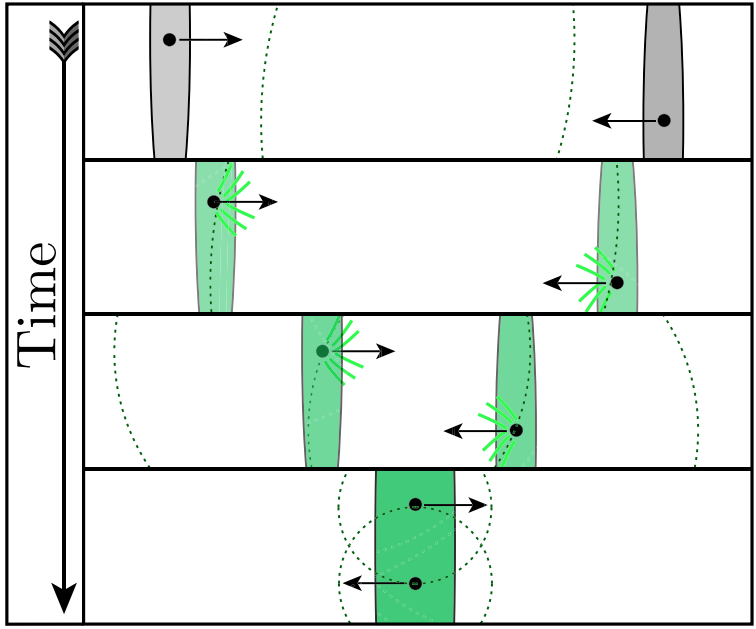}
  \caption{\small Stages of ultrarelativistic scattering.
  The dotted arcs indicate the separations at which closed string radiation becomes relevant.  As the branes approach each other, the rapidity decreases and the radii of the circles decrease.  The green pancakes represent the D-branes and their bremsstrahlung; the latter accounts for most of the initial energy, and does not substantially outpace the corresponding source by $\tau=0$.}\label{scatteringprocess}
\end{figure}

\begin{figure}[h]
	\centering
		\includegraphics[width=0.4\textwidth]{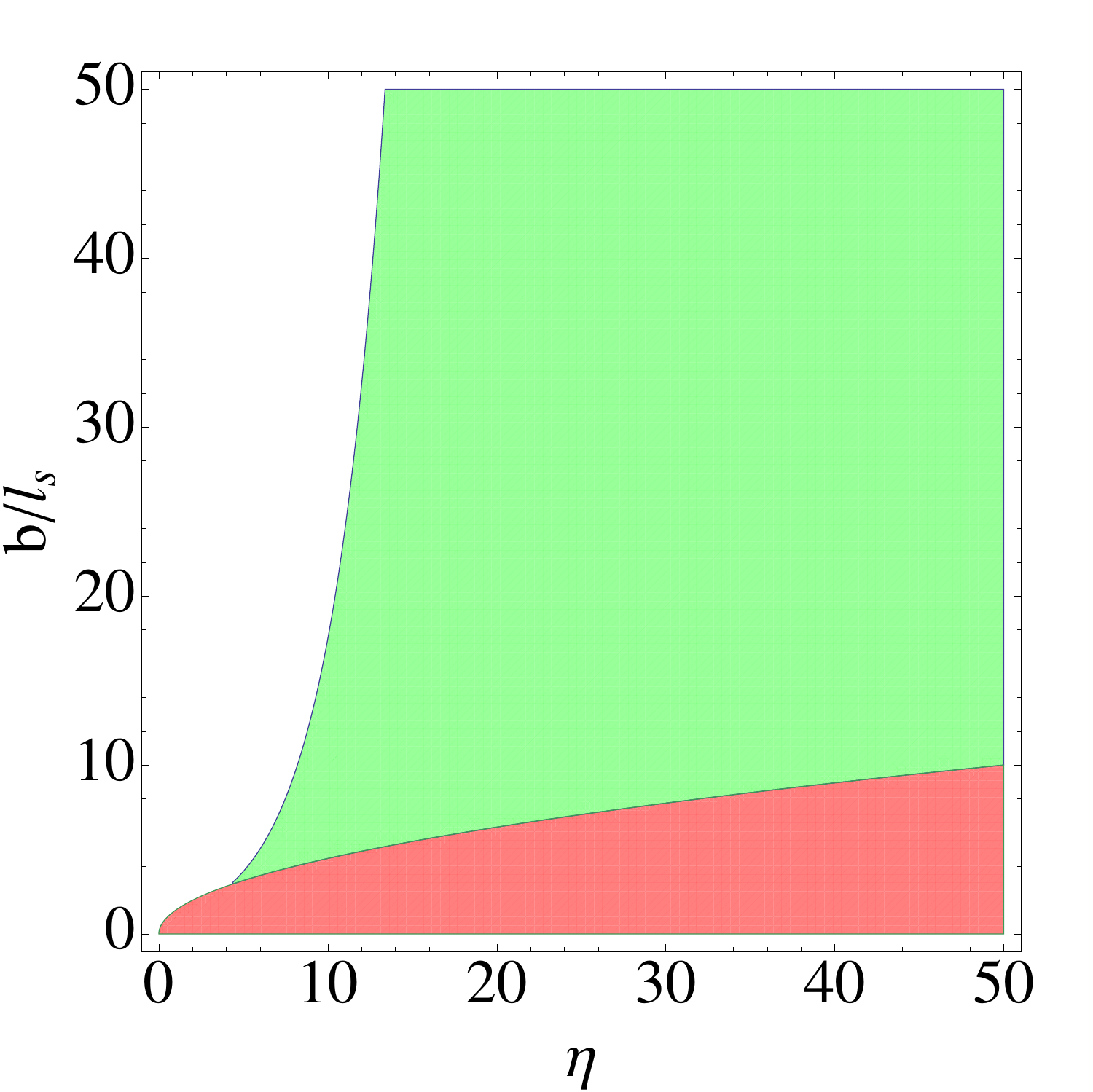}
	\caption{\small Regions of the parameter space. In the white region, the energy lost to bremsstrahlung is insignificant.
In the green region,  bremsstrahlung is important and the approximations made in our analysis are consistent.  In the red region, the
closed string exchange potential, \eq{relpot}, breaks down.  The case $g_{s}=0.1$ and $k=3$ is shown.}\label{parameterspace}
\end{figure}

The fact that a large amount of energy is radiated at $\tau_{\text{brem}}$ does not imply that the branes decelerate to nonrelativistic motion immediately.
Instead, rapid losses due to bremsstrahlung quickly decelerate the D-brane to the critical rapidity
\be
\eta(\tau)\sim {1\over 2k-15}\text{log}\left[{(8-k)(9-k)-1\over 2\sigma(k)}  G_{10}^{-1}T_{\te{D}k}^{7-k} b^{k-8} |\tau| ^{(8-k) (9-k)-1} \right]\,,\label{eta}
\ee for which the energy radiated in a given time interval is of order the D-brane kinetic energy at the beginning of this interval.
Thus, the rapidity $\eta$ decreases as
\be
\dot{\eta}(\tau) \sim \frac{(8-k) (9-k)-1}{15-2k} \cdot \frac{1}{\tau} \,.\label{eta1}
\ee

We are now able to estimate the critical impact parameter $b_{\te{int}}$ above which the scattering process does not lead to a significant loss of kinetic energy in bremsstrahlung. To obtain an upper limit on the radiated energy density we note that $a_{y}\lesssim a_{r}={ {\partial V(r)\over \partial r}/ \rho_{\te{D}k}(\tau)}$, where the bound is saturated in the limit of large impact parameters. Using \eq{closedproduc} we define $b_{\te{int}}$ by
\be \label{interactiondistance}
\int _{-\infty}^{\infty}d\tau~{G_{10} T^{2}_{\te{D}k}} \cdot\left(\gamma^{2}a_{r}(b=b_{\te{int}})\right)^{8-k} \equiv \gamma \,T_{\te{D}k}\,,
\ee
from which we find
\be \label{bintis}
b_{\te{int}} \propto \exp\bl{(15-2k )\eta\over (8-k)(9-k)-1 }\br \,,
\ee which is exponentially large in the ultrarelativistic limit.   Combining \eq{bintis} and \eq{closedstop}, we conclude that ultrarelativistic D-branes experience rapid energy loss to bremsstrahlung even when their transverse and longitudinal separations are very large in string units.

Thus far we have neglected energy lost to open string pair production, emphasizing the critical role of bremsstrahlung.
This is a consistent approximation: for large initial rapidity $\eta_0$, bremsstrahlung becomes significant at distances $\tau_{\te{brem}}$ that are exponentially large in string units, and
open string pair production is entirely negligible at such large separations: see Appendix \ref{sec:onlyopenstrings}.
Thus, closed string radiation dramatically decelerates the D-brane pair long before open string pair production becomes significant.\footnote{It would be interesting to understand whether  bremsstrahlung affects the closely related process of flux discharge by pair production, as described in Refs.~\cite{Kleban,D'Amico:2012ji}.}

We have numerically evolved the system including the effects of gravitational bremsstrahlung, finding excellent agreement with the analytic result \eq{eta}, as shown in Figure \ref{short1} for the parameter choice $\eta_{0}=26$ and $b=10\,l_{\te{s}}$.

A further consideration is the possibility that, for impact parameters $b$ smaller than some critical value $b_{\te{BH}}(\eta)$, the final state of the scattering D-brane pair is a black hole.
We will not obtain a complete description of the late-time dynamics, but will estimate $b_{\te{BH}}(\eta)$ using the cross section for black hole formation in high-energy particle scattering. The cross section in the center of mass frame is given by $\sigma_{\te{BH}}\sim \pi (R_\te{s}^{\perp})^2$ \cite{Banks:1999gd,Eardley:2002re}, where $R_\te{s}^{\perp}$ is the Schwarzschild radius associated to the total energy $E$ in the center of mass frame, $E \approx 2 T_{\te{D}k} \gamma_\te{COM} = T_{\te{D}k} e^{\eta_0/2}$.  Explicitly, we have \cite{Horowitz}
\be
 \left({R_{s}^{\perp}\over{
l_{s}}}\right)^{7-k} = g_s {\left({{7-k}\over{8-k}}\right)}
(2\pi)^{7-k}  s_{k}\cdot \gamma_\te{COM}\,,  \label{rsperp}
\ee
where $s_k=2^{5-k} \pi^{(5-k)/2} \Gamma((7-k)/2)$.
Thus, for $b \lesssim R_{s}^{\perp}(\eta_0)$, the final state is plausibly a  black hole.

One might object that bremsstrahlung dramatically reduces the kinetic energy of each of the branes, and hence reduces $R_{s}^{\perp}$.
However, most of the bremsstrahlung is very forward, and is contained within a cone of opening angle $\theta_{\te{c}}\sim1/\gamma_\te{COM}$.
Moreover, the bremsstrahlung does not substantially outpace the D-brane that emitted it: at $\tau=0$, the longitudinal distance from the horizon of the moving brane to the leading edge of the radiation it emitted at $\tau=\tau_{\te{brem}}$ is parametrically smaller than $R_{s}^{\perp}$.
For $b < b_{\te{int}}$ and $\eta_0 \gg 1$, one finds from \eq{closedstop1} and \eq{rsperp} that the majority of the energy radiated forms a thin pancake of transverse size $\sim R_{s}^{\perp}$ at $\tau=0$.  Thus, an order-unity fraction of the initial energy of the system fits inside a region of maximum diameter $\sim R_{s}^{\perp}$.

We summarize the stages of the scattering process as seen from the center of mass frame in Figure~\ref{scatteringprocess}, and in Figure~\ref{parameterspace} we display the regions of the parameter space in which our approximations are applicable.

\section{Dissipation in DBI Inflation}\label{sec:applications}

We have shown above that even a small acceleration of an ultrarelativistic D-brane leads to efficient energy loss through bremsstrahlung. This effect may alter the dynamics of inflationary models involving relativistic D-branes, most notably DBI inflation \cite{Silverstein:2003hf,Alishahiha:2004eh}.  In this section we will consider the consequences of bremsstrahlung for the DBI model.

The best-understood realization of DBI inflation involves a D3-brane in a warped throat region of a compactification, and for concreteness we will restrict our attention to this scenario.  The idea is that a strong potential impels the D3-brane to move radially inward, toward the tip of the throat, at a speed that approaches the local limiting speed.  The result is a period of accelerated expansion, with distinctive signatures in the primordial perturbations.

A D3-brane in a compactification with stabilized moduli experiences a potential induced by Planck-suppressed couplings to the moduli-stabilizing energy sources.
For a D3-brane in a warped throat region --- for definiteness, let us consider a Klebanov-Strassler throat \cite{KS} --- the structure of this potential is well-understood \cite{Baumann:2010sx}.  The potential is a function of all six coordinates of the throat, and in particular the potential in the angular directions is nonvanishing.\footnote{This can be understood simply on symmetry grounds: the compact bulk cannot respect any exact continuous isometries, and the corresponding lifting of the isometries that are present in the noncompact limit leads to angular gradients of the supergravity fields in the throat region.}  The characteristic scale of the masses in the radial and angular directions is the Hubble scale $H$.  For successful DBI inflation, one needs a vastly larger {\it{radial}} mass, to drive the D3-brane to relativistic speeds.  Although such a mass seems hard to obtain \cite{Agarwal:2011wm} in the moduli potential computed in Ref.~\cite{Baumann:2010sx}, for the present analysis we will grant the existence of a suitable potential in the radial direction.

In light of our results so far, there are two natural questions concerning bremsstrahlung in the DBI scenario.
First, do the angular forces resulting from a natural moduli potential necessarily cause the infalling D3-brane to accelerate transverse to its radial trajectory, and thus to lose a large fraction of its energy in bremsstrahlung?
Second, does the purely radial potential itself lead to significant bremsstrahlung due to longitudinal acceleration?

To develop intuition for this problem we observe that there is a very well known electromagnetic analogy: a charged particle accelerated along a curving trajectory by an electromagnetic field, e.g.\
in a synchrotron, reaches a maximum velocity when the rate of energy loss in synchrotron radiation balances the energy input from the electromagnetic field.  In the DBI example, the curving trajectory is a consequence of
nonvanishing angular gradients of the moduli potential, which as we explained above are
unavoidable in stabilized compactifications.  Although the essential physical question is the same,
the DBI problem has important differences from the electromagnetic problem: the object in motion has spatial extent;
the four-dimensional spacetime is expanding; and the radiation
spreads\footnote{One can check that even the longest wavelengths used in our estimates  of the total radiated power are much smaller than the radius of the extra dimensions.} in a six-dimensional compact space.

The significance of bremsstrahlung from either transverse or parallel acceleration can be assessed by determining the limiting speed set by radiative losses, or equivalently the limiting $\gamma$ factor $\gamma_{\te{brem}}$, and comparing this to the $\gamma$ factor $\gamma_{\te{DBI}}$ that is obtained by neglecting radiation.  When $\gamma_{\te{brem}} \lesssim \gamma_{\te{DBI}}$, bremsstrahlung sets the most stringent upper limit on the speed, and radiative losses control the dynamics.  We now consider the transverse and parallel  contributions in turn.

\subsection{Radiation from transverse acceleration}

We will write the D3-brane potential in the form  \cite{Baumann:2010sx}
\be
V(r,\Psi)= \frac{m_{DBI}^2}{2}\phi^2 + \mu^{4} \sum_{Q}c_{Q}(\phi/\phi_{\te{UV}})^{\Delta(Q)}f_{Q}(\Psi)\, ,
\ee
where $\phi=\sqrt{T_{\te{D}3}}r$ is the canonically normalized radial field, with $r$ the radial coordinate; $\phi_{\te{UV}}$ is the location of the top of the throat; and $\Psi$ parameterizes the five angles.

The index $Q$ represents possible quantum numbers under the angular isometries; $\Delta(Q)$ is then related to the corresponding operator dimension, $f_{Q}(\Psi)$ is an order-unity function of the angles, and $c_Q$ is the associated Wilson coefficient.  The first term, $m_{DBI}^2\phi^2/2$, is a significant mass term that controls the radial evolution.  The remaining terms are contributions from sources in the bulk of the compactification, and the scale $\mu^{4}$ measures the significance of these effects.

We will consider only a single bulk contribution with $\Delta=2$,\footnote{In the case of the  Klebanov-Strassler geometry, there are terms with $\Delta=1$ and $\Delta=3/2$ that correspond to parametrically stronger angular forces, but for a conservative, general, and analytically simple estimate we take $\Delta=2$.}
 writing
\be \label{simplemodel}
V(\phi,\hat\psi)= \frac{1}{2}{m_{DBI}^2}\phi^2 + \frac{1}{2}{m_{{\cal B}}^2} \hat{\psi}^2\,,
\ee where $\hat{\psi} = \phi \,\psi$ is the canonically-normalized field corresponding to the  dimensionless angular coordinate $\psi$, and $m_{{\cal B}} \sim \mu^2/\phi_{\te{UV}}$ sets the scale of the bulk effects.

As demonstrated in Ref.~\cite{Silverstein:2003hf}, the DBI model with a quadratic potential of the form $V(\phi)= \frac{1}{2} m_{DBI}^2\phi^2$ leads to prolonged inflation if $m_{DBI}^2 \gtrsim g_{\te{s}}M_p^{2}/\lambda$, where  $\lambda$ is the 't Hooft coupling.
The total number of e-folds that result in such a potential is  \cite{Alishahiha:2004eh}
\be
N_{e}=\sqrt{\lambda m_{DBI}^2\over 6 g_{\te{s}} M_p^2}\, .\label{efolds}
\ee

We assume that the masses $m_{{\cal B}}$ induced by compactification have the natural scale $m_{{\cal B}} \sim H \ll m_{DBI}$, where $H$ is the Hubble parameter. (This  property has been checked in a wide range of configurations and is explained in detail in Ref.~\cite{Baumann:2010sx}.)  Thus, the angular motion is not heavily  damped  by Hubble friction, and the angular acceleration is approximately
\begin{equation} \label{ais}
a \approx \frac{m_{{\cal B}}^2\, \phi}{\sqrt{T_{\te{D}3}}\gamma}\,.
\end{equation}

The reader may wonder why the D3-brane does not slide to the minimum of the angular potential and then proceed radially  inward without undergoing further angular acceleration.
The point is that there are many contributions to the potential that scale differently with the radial coordinate
$\phi$, so that the location of the instantaneous minimum in the angular directions depends on $\phi$.  Thus, the simple model \eq{simplemodel} describes the angular potential for a small range of $\phi$, and --- unless $m_{{\cal B}} \gtrsim m_{DBI}$, which is uncommon --- the D3-brane does not track the instantaneous minimum of the angular potential.  The resulting picture is rather similar to a wiggler in a synchrotron.

The rate at which bremsstrahlung dissipates energy, for a given Lorentz factor and transverse acceleration $a$, is given by \eq{closedproduc} as
\begin{equation} \label{pbrem}
P_{\te{brem}} = G_{10} T_{\te{D}3}^2 \gamma^{10} a^{5} \,.
\end{equation}

For a D3-brane experiencing a quadratic potential in a warped throat, the Lorentz factor $\gamma_{\te{DBI}}$ is given, for $N_e \gg 1$, by \cite{Silverstein:2003hf,Alishahiha:2004eh}
\be
\gamma^{2}_{\te{DBI}} \approx {4\over (2\pi)^{3}}N_e^2 \left(\frac{M_p}{\phi}\right)^4\,.\label{dbigammasimple}
\ee
The maximum rate at which the radial potential can supply kinetic energy is
\begin{equation}
P_{\te{infall}} = \dot\phi\, \partial_{\phi} V\,,
\end{equation} while the speed limit set by the DBI kinetic term is
\begin{equation}
\dot\phi_{\te{max}} = \frac{\phi^2}{\sqrt{\lambda}} g_s^{1/2}(2\pi)^{3/2}\,,
\end{equation}
where we use the conventions of Ref.~\cite{Baumann:2006cd} for the DBI Lagrangian.  Thus,
\begin{equation} \label{pinfall}
P_{\te{infall}} = \frac{m_{DBI}^2g_{s}^{1/2} (2\pi)^{3/2}}{ \sqrt{\lambda}}\phi^3\,.
\end{equation}

We would like to understand whether, for some values of the parameters, $P_{\te{brem}}$ is comparable to $P_{\te{infall}}$, so that losses to bremsstrahlung can equal or exceed the rate of energy input from the gradient of the radial potential.  For this purpose, it will prove useful to incorporate the COBE normalization, in order to specify the inflationary Hubble scale $H$ in terms of the remaining parameters.\footnote{Our goal is to  give a parametric description of dissipation in the DBI scenario, not to specify choices of microphysical parameters that are consistent with all observations.  Nevertheless, it is sensible to use the COBE normalization in order to set the inflationary scale in the general range that is appropriate for  a large-field model.}  One can show that (see e.g.\ Ref.~\cite{Baumann:2006cd})
\begin{equation} \label{hovermp}
\frac{H}{M_p} =\frac{4\pi}{\gamma\,\phi}\cdot\sqrt{P_{{\cal{R}}}}\,,
\end{equation} where $P_{\cal{R}} = \frac{25}{4}\cdot(1.91\cdot 10^{-5})^2$.

Now, setting $m_{{\cal B}} \approx H$ and using \eq{ais} and \eq{hovermp} in \eq{pbrem}, and combining this result with \eq{pinfall}, \eq{dbigammasimple}, and \eq{efolds}, we find that
$P_{\te{brem}}(\phi) \approx P_{\te{infall}}(\phi)$ for $\phi \lesssim \phi_{\star}$, where
\begin{equation}
\phi_{\star} = 4\cdot 10^{11} \cdot \frac{N_e^{7/2}}{g_s^{1/2} \alpha'^{5/2} M_p^{4} \lambda^{3/4}}\, .\label{phicross}
\end{equation}
Because $P_{\te{brem,trans}} \propto \phi^5$ and $P_{\te{infall}} \propto \phi^3$, the dynamics is controlled by bremsstrahlung for $\phi \gtrsim \phi_{\star}$, while for $\phi \lesssim \phi_{\star}$ bremsstrahlung  can be neglected.  Because of the large prefactor, ultimately arising from the COBE normalization \eq{hovermp}, we easily see  that \eq{phicross} does not present a meaningful constraint on  the DBI  scenario.   In short,  angular forces of the natural magnitude do  lead to dissipation through radiation from transverse acceleration,  but this dissipation is generally small compared to the energy supplied by the steep radial potential.

\subsection{Radiation from longitudinal acceleration}

Having seen that angular forces of the natural scale expected in a stabilized compactification do not lead to dramatic losses  due to transverse acceleration, we now turn to a very similar analysis for losses due to longitudinal acceleration.   At first sight one might  expect  radiation from transverse acceleration to provide the dominant loss mechanism,  and this  expectation would be correct if the radial  and angular forces were comparable.
However, in the DBI  scenario the radial potential is necessarily rather steep compared to the natural scale (which is the Hubble scale $H$), and in fact we will find that longitudinal acceleration leads to substantial losses.

We now have a purely quadratic potential,
\be
V(\phi)= \frac{1}{2}{m_{DBI}^2}\phi^2 \,,
\ee  leading to the acceleration
\begin{equation}
a \approx \frac{m_{{DBI}}^2\, \phi}{\sqrt{T_{\te{D}3}}\gamma^3}\,.
\end{equation}
Using our result \eq{powerdens}  for the power radiated\footnote{Recall that \eq{powerdens}  is a conservative estimate that includes only that portion of the radiated power that is well-described by classical emission.} in longitudinal acceleration, we have
\begin{equation} \label{longitudepbrem}
P_{\te{brem}} =G_{10} T_{\te{D}3}^{3/2} \gamma^{45/2} a^{7}  \,.
\end{equation}
As in the case of transverse acceleration, $P_{\te{brem,long.}} \propto \phi^5$ and $P_{\te{infall}} \propto \phi^3$, but now the overall magnitude of the radiated power is larger: we find that $P_{\te{brem,long.}} \approx P_{\te{infall}}$ at
\begin{equation}
\phi_{\star} \approx 2\cdot 10^{-8} \cdot   \frac{\lambda^{11/2}}{\pi^{37/4} g_s^{19/2} \alpha'^{8} M_p^{15} N_e^{27/2}}\,.  \label{longitudinalDBI}
\end{equation}
Radiation from longitudinal acceleration controls the dynamics for $\phi \gtrsim \phi_{\star}$.

We conclude that a systematic treatment of bremsstrahlung  is a prerequisite for a reliable computation of the trajectory at $\phi \gtrsim \phi_{\star}$.
On the other hand, neglecting bremsstrahlung is not necessarily consistent in the complementary regime $\phi \lesssim \phi_{\star}$: in obtaining \eq{longitudinalDBI}
we did not solve the fully backreacted equations of motion, and our estimates of radiated  power were conservative throughout.

\subsection{Constraints from bremsstrahlung}

Thus far our calculation has not depended on specific properties of the compactification: we have simply assumed that the radial and angular potentials are quadratic, and used the natural relation $m_{{\cal B}} \approx H$.
However, it is very instructive to impose the microphysical constraints that arise in a warped throat region: in particular, the maximum value $\phi_{\te{UV}}$ of the $\phi$ coordinate can be related to the volume of the compactification.
Following Ref.~\cite{Baumann:2006cd}, we write the total warped volume ${\cal V} \alpha^{\prime 3}$ of the compactification as
\begin{equation}
{\cal V} \alpha^{\prime 3} = (V_{6}^{w})_{\te{throat}} + (V_{6}^{w})_{\te{bulk}} \equiv (1+\sigma) (V_{6}^{w})_{\te{throat}} = (1+\sigma) \frac{N}{4}G_{10}\,\phi_{\te{UV}}^2\,,
\end{equation} where $N$ is the D3-brane charge of the throat.
Moreover, we have the usual relation $M_p^2 = {\cal V} \alpha^{\prime 3}/G_{10}$, with
\begin{equation} \label{gtenis}
G_{10} = \frac{1}{2}(2\pi)^7 g_s^2 \alpha^{\prime 4}\,,
\end{equation} so that
\begin{equation}
M_p^2 = (1+\sigma) \frac{N}{4}\phi_{\te{UV}}^2 \,.
\end{equation}
Finally, in a Klebanov-Strassler throat, we have $\lambda=\frac{27}{4}\pi g_{s}N$.
Combining the above results, the critical location $\phi_{\star}$ can be expressed in the form

\begin{equation}
\frac{\phi_{\star}}{\phi_{\te{UV}}} \approx  10^{37} \cdot \frac{g_s^{12}
N^{6}(1+\sigma)^{1/2}}{N_e^{27/2}{\cal V}^{8}}  \, ,\label{phicrossuv}
\end{equation}
and we remind the reader that $\frac{\phi_{\star}}{\phi_{\te{UV}}}$ corresponds to the warp factor at the location $\phi=\phi_{\star}$.
As an example, taking the conservative values\footnote{Evidently, at weak coupling  and large volume, holding $N$  fixed, the  impact of bremsstrahlung is  dramatically increased.}
$g_s = 0.5, N_e = 60, {\cal V} = 1000, N=100$, and $\sigma=10$, we find $\phi_{\star}/\phi_{\te{UV}} \approx 8\cdot 10^{-3}$.
In this case the warp factor decreases by about two orders of magnitude before the  (instantaneous) effect of bremsstrahlung
can be neglected as a correction to the dynamics.  Moreover, even for $\phi \ll \phi_{\star}$  the  actual trajectory will differ
from that computed  by neglecting dissipation: the velocity at $\phi_{\star}$ is diminished  by the history of dissipation.

We conclude that for DBI inflation driven by a quadratic potential in a warped throat region, bremsstrahlung can provide a significant dissipative effect.
In some parameter regimes this presents a strong constraint, while in others it is self-consistent to neglect radiative losses.

\section{Conclusions}\label{sec:conclusions}

We have shown that in the ultrarelativistic limit, D-brane scattering involves extremely rapid dissipation via bremsstrahlung of massless closed strings.  The initial trigger for this process is the velocity-dependent force arising from closed string exchange.  Except in special cases, the density of open string pairs remains small, with nearly all the initial energy lost to bremsstrahlung.

In the course of our analysis, we characterized radiation from a $k$-dimensional extended object undergoing relativistic accelerated motion in an even-dimensional background spacetime.
We obtained a lower bound on the rate of energy loss in this system due to bremsstrahlung, and we determined the critical Hagedorn acceleration beyond which the power radiated from a D-brane diverges due to emission of massive closed strings.

The intense dissipation experienced by ultrarelativistic D-branes has implications for cosmological models involving D-brane motion.  We have shown that bremsstrahlung provides a  drag force in DBI inflation \cite{Silverstein:2003hf,Alishahiha:2004eh}, and  that in reasonable compactifications this effect can significantly alter the cosmological evolution.  It would be  worthwhile to understand the implications of our results for models involving repeated interactions, such as trapped inflation \cite{Green:2009ds} or unwinding inflation \cite{D'Amico:2012ji},  which yield novel contributions to the primordial scalar and tensor perturbations (see also Refs.~\cite{Romano:2008rr,Brax:2010tq,Barnaby:2010sq,Senatore:2011sp}).  When the scattering events are relativistic,  bremsstrahlung will very plausibly increase the rate of energy loss, and could lead to modified signatures.

\subsection*{Acknowledgements}
We are grateful to D.~Baumann, D.~Chernoff, A.~Dymarsky, E.~Flanagan, M.~Kleban, M.~M.~Sheikh-Jabbari, E.~Silverstein, S.~Teukolsky, and H.~Tye for helpful discussions.
The research of L.M. was supported by the Alfred P. Sloan Foundation, by an NSF CAREER Award, and by the NSF under grant PHY-0757868. The work of T.B. was partially supported by a Cornell Graduate School Fellowship and by the NSF under grant PHY-0757868.

\appendix
\section{Energy Loss from Open String Pair Creation}\label{sec:onlyopenstrings}
In this appendix we review the production of open strings in relativistic D-brane scattering, following Ref.~\cite{McAllister:2004gd}.
Considering D-branes on fixed trajectories and neglecting losses to bremsstrahlung, we obtain the pair production rate and the time $\tau_{\te{open}}$ at which the energy stored in open string pairs is comparable to the initial kinetic energy.  As $|\tau_{\te{open}}| \ll |\tau_{\te{brem}}|$, bremsstrahlung takes effect at parametrically larger distances than those relevant for open string pair creation, so that it is consistent to neglect open string pair production for most of the scattering process, as we have done in the main text.  Even so, open string pairs may become significant near the time of closest approach, for certain initial conditions.

We stress that neglecting closed string radiation is not generally a consistent approximation.  With radiation  losses turned off, relativistic scattering induces exponentially large densities of open strings \cite{McAllister:2004gd}, but on trajectories decelerated by bremsstrahlung, radiation of closed strings can absorb energy so quickly that a dense population of excited open strings never arises.
Nevertheless, focusing on open strings alone allows us to place an upper bound on the rate of energy loss to open strings, which is required to establish the consistency of the approximation of neglecting open string pair production that was made in \S\ref{sec:openandclosed}.

We begin by examining the imaginary part of the amplitude \eq{ampA}, following Refs.~\cite{Bachas:1992bh,Bachas:1995kx,McAllister:2004gd}.
First, we evaluate the integral over the complex $t$ plane \cite{Bachas:1995kx}:
\bea \label{annulusresult}
\text{Im}({\cal A})&=&2\pi\sum_{j}\te{Res}_{j\in \mathbb{N}}\left[{V_{\te{D}p}  \over j} \left({8 \pi^3 \alpha'  j\over \eta}\right)^{-p/2} \exp \left(-{{ j b^2} \over 2\eta\alpha'}\right) {\theta_{11}(j/2, i j \pi/\eta)^4 \over {\theta_{11}( j, i j \pi/\eta)\eta_{\te{D}}(i j \pi/\eta)^{9}}}\right]\nonumber\\
&=&{V_{\te{D}p} \over {(2 \pi)^p}} \sum_{j=1,3,...}^{\infty} {1 \over j}
 \left(\eta \over{2 \pi j \alpha^{\prime}} \right)^{p/2} \exp\Bigl(-{{b^2 j}\over{2 \alpha^{\prime}\eta}}\Bigr){\theta_{10}^4\left(0,i{\pi j\over \eta}\right)\over \eta_{\te{D}}^{12}\left(i{\pi j\over \eta}\right)}\,.  \label{impart}
\eea
Recognizing the final ratio as the partition function $Z$ as in Ref.~\cite{McAllister:2004gd},
\begin{equation}
  Z\Bigl(i\frac{j\pi}{\eta}\Bigr)={\theta_{10}^4\left(0,i{\pi j\over \eta}\right)\over \eta_{\te{D}}^{12}\left(i{\pi j\over \eta}\right)} = \sum_{n=0}^{\infty}N(n)\,{\rm{exp}} \Bigl(-\frac{2\pi^2 n j}{\eta}\Bigr)  \,,
\end{equation} we can write $\text{Im}({\cal A})$ in the suggestive form
\begin{equation} \label{Schwingerlike}
\text{Im}({\cal A})= { V_{\te{D}p} \over {(2 \pi)^p}} \sum_{j=1,3,...}^{\infty} {1 \over j} \left(\eta \over{2\pi j \alpha'} \right)^{p/2} \sum_{n=0}^{\infty}N(n)\,
\exp\biggl(-\frac{2\pi^2}{\eta}\,j\,\left[\frac{b^2 }{4\pi^2  \alpha^{\prime}}+n \right]\biggr)\,.
\end{equation}
This expression is reminiscent of Schwinger's famous formula \cite{Schwinger} for the rate at which electron pairs are produced in a constant electric field, with two novelties: \eq{Schwingerlike} contains a sum over  the string level $n$, weighted by the degeneracy $N(n)$; and the factor of
$\eta$ in the exponential is suggestive of relativistic corrections to the mass of the produced pairs.
To make this point precise, and to obtain the instantaneous rate\footnote{One might be tempted to state that by the optical theorem, the rate of production of open strings stretched between the two branes is related to
$\text{Im}({\cal A})$.  In fact, because of the Wick rotation relating branes in motion to branes at a complex relative angle, $\text{Im}({\cal A})$ is related to the total number of strings produced in the scattering process, rather than to the rate of production.} of pair production of massive open strings, we will relate the total pair production derived from the annulus calculation,  \eq{annulusresult},  to the corresponding result in quantum field theory, where the instantaneous pair production rate can be computed by standard methods.
Thus, the idea is to describe open string pair production in the field theory arising from level truncation.  The primary subtlety that arises is that the mass spectrum of open strings stretched between D-branes in relativistic relative motion is rescaled in comparison to the spectrum of open strings stretching between D-branes at rest.  We now proceed to a careful treatment of this point, building on and extending the results of Ref.~\cite{McAllister:2004gd}.

Consider the field theory described by the Lagrange density
\be \label{QFTL}
\mathcal{L} ={1\over 2} \partial_{\mu}\phi \partial^{\mu}\bar{\phi}+{1\over 2} \partial_{\mu}\chi \partial^{\mu}\bar{\chi}- \frac{1}{2} m_{\chi}^{2}  \chi\bar{\chi}\,.
\ee
A mode $u_{k}$ with momentum $k$ obeys the wave equation
\be
\left[ \partial_{t}^{2}+k^{2}+m_{\chi}^{2} \right]u_{k}=0\,.
\ee
To represent the string spectrum, we consider a collection of fields $\chi_n$, with
\begin{equation}
m_{\chi_n}=\left({|\phi|^{2}\over 4\pi^{2}\alpha^{\prime 2}} +{n\over \alpha^{\prime} }\right)\,,
\end{equation}
where $\phi(t)=i b+v t$, and we take the number of fields $\chi_n$ at level $n$ to be $N(n)$.  We will take the $\chi_n$ to interact much more strongly with $\phi$ than with each other, so that the pair production rates of each species are simply additive.  For any  finite maximum level $n_{max}$, this quantum field theory corresponds to the level-truncated string field theory whose mass spectrum  is that of open strings between D$p$-branes {\it{at rest}} at separation $b$.

With some foresight, we now introduce the rescaling
\be \label{rescaling}
k^{2}+m_{\chi_n}^2\rightarrow {v^{2}\over \eta^{2}}\left(k^{2}+m_{\chi_n}^2\right)\, .
\ee
We can now use the WKB approximation to find an approximate solution to the rescaled wave equation, for a given $n$:
\be
u_{k}^{in}\propto \exp\left[ -i \int_{t} {v\over \eta}\sqrt{k^{2}+\left({b^{2}+v^{2}t^{\prime2}\over 4\pi^{2}\alpha^{\prime 2}} +{n\over \alpha^{\prime} }\right)}\, dt^{\prime} \right]\,,
\ee
where the integral is taken over a large semicircle below the real axis in the complex $t^{\prime}$ plane. The integral in the exponential determines the ratio of the reflected and transmitted amplitudes, corresponding to the particle density $n_{k}$. At large $t$ we can expand,
\be
{v\over \eta}\sqrt{k^{2}+{b^{2}+v^{2}t^{\prime2}\over 4\pi^{2}\alpha^{\prime 2}} +{n\over \alpha^{\prime} }} \sim \frac{v^2t^{\prime}}{2 \pi \eta \alpha^{\prime}}
+\frac{\pi}{\eta t^{\prime}}\Bigl({k^{2} \alpha^{\prime}+{b^{2}\over 4\pi^{2}\alpha^{\prime}} + n}\Bigr)\,.
\ee
For a single species, this gives (see Ref.~\cite{Kofman:2004yc} for a more detailed treatment of this point)
\be
n_{k}=\exp\left(-{2\pi^{2}\over \eta}\left[k^{2}\alpha^{\prime}+{b^{2}\over4\pi^{2}\alpha^{\prime}}+n \right] \right)\,.
\ee
Including the level degeneracy $N(n)$, the total number of particles in $p$ dimensions is thus
\be
N_{\te{tot}} =\sum_{n=0}^{\infty}N(n)\, {1\over (2\pi)^{p}}\int d^{p}k ~n_{k}={1\over (2\pi)^{p}}\left( {\eta\over 2\pi \alpha^{\prime}}\right)^{p/2}\sum_{n=0}^{\infty}N(n)\,
\exp\left(-{2\pi^{2}\over \eta}\left[{b^{2}\over4\pi^{2}\alpha^{\prime}}+n \right] \right)\,.
\ee
This field-theoretic result agrees precisely with the $j=1$ term of the string theory result \eq{Schwingerlike}.
The terms at higher $j$ in \eq{Schwingerlike} correspond to multi-instanton effects (see e.g.\ Ref.~\cite{pagekim}), and can likewise be obtained in field theory, though we will not do so here.

We therefore conclude that the result of \eq{Schwingerlike} for open string pair production in relativistic D-brane scattering is simply described by summing field-theoretic results over the accessible species of open strings, {\it{provided}} that we use correct spectrum of string states.  Namely, the rate given in \eq{Schwingerlike} is precisely reproduced in a field theory in which the energies of excited open strings
are suppressed according to the rescaling of the effective energy\footnote{This result is stated incorrectly in the JHEP version of Ref.~\cite{McAllister:2004gd}, but is stated correctly in the arXiv version.
We thank M.~Kleban for drawing our attention to this discrepancy.}
\be \label{tensionrescaling}
E(n)^2 = \frac{v^2}{\eta^2}\Biggl(\frac{n}{\alpha^{\prime}} + \frac{b^2+v^2\tau^2}{4\pi^2{\alpha^{\prime}}^2} \Biggr)\,,
\ee
so that pair production of highly-excited strings is substantially enhanced compared to a naive computation based on the spectrum of strings stretched between branes at rest \cite{McAllister:2004gd}.
The rescaling \eq{tensionrescaling} can also be understood in the T-dual picture of open strings in an electric field, as reviewed in Ref.~\cite{Ambjorn:2000yr}.

We remark that the energy given in \eq{tensionrescaling}
corresponds to the energy of an open string as measured in the fixed target frame in which one of the branes is at rest, just as the time variable $\tau$ measures the time in that frame --- this can be checked in the T-dual electric field picture.  This leads to a correction to the stopping length computed in Ref.~\cite{McAllister:2004gd}, but the main conclusions of Ref.~\cite{McAllister:2004gd} are unchanged.

Equipped with the total amount of string production, we now turn to determining the time-dependent production rate that is consistent with the total string production \eq{impart} and with the non-adiabaticity $\xi \equiv \dot{\omega}/\omega^{2}$.
The idea is to write a physically-motivated model of the time-dependence and verify numerically, in the quantum field theory case where time-dependent production can be computed directly, that the model gives dynamics in good agreement with the full treatment.
Following Ref.~\cite{McAllister:2004gd}, we  write
\bea \label{productionrate}
\text{Im}({\cal{A}} )&=&\int_{-\infty}^{\infty}d\tau~\zeta \xi^2(\tau){v\over 2\pi} {V_{\te{D}p} \over {(2 \pi)^p}} \sum_{j=1,3,...}^{\infty} {1 \over j} \left(\eta \over{2\pi j} \right)^{(p-1)/2} \exp\Bigl(-{{(b^2+v^{2}\tau^{2}) j}\over{2 \eta}}\Bigr){\theta_{10}^4\left(0,i{\pi j\over \eta}\right)\over \eta_{\te{D}}^{12}\left(0,i{\pi j\over \eta}\right)}\nonumber\\
&\equiv&\int_{-\infty}^{\infty}d\tau~ \Gamma_{\text{prod}}(\tau)\,,
\eea
where a Gaussian dependence on the separation and a quadratic dependence on the non-adiabaticity parameter $\xi=\dot\omega/\omega^2$ were introduced. The normalization constant $\zeta$ can be determined by stipulating that the total amount of open string production agrees with the annulus calculation.

\begin{figure}[t]
	\centering
		\includegraphics[width=0.7\textwidth]{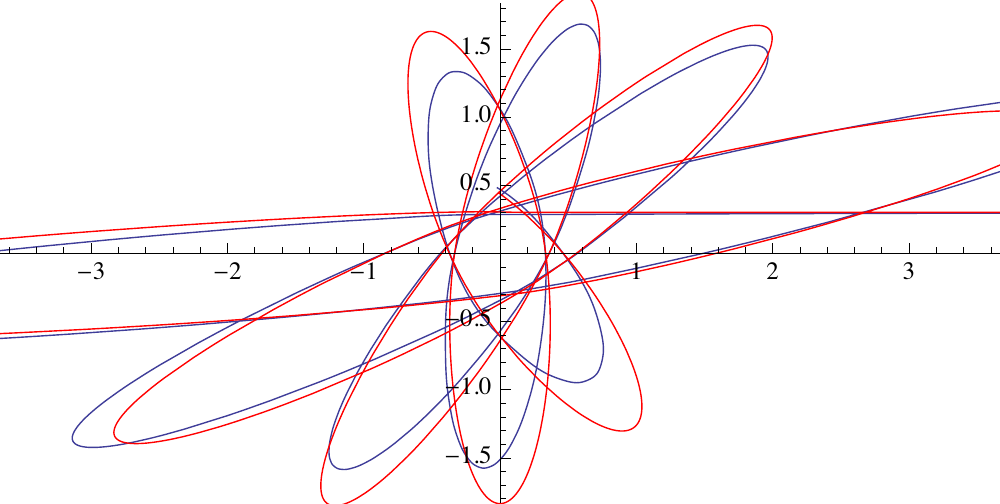}
	\caption{\small Plot of the D-brane trajectories for the initial conditions $v_{0}=-1$, $x_{0}=10$, $y_{0}=0.3$, $g=\sqrt{20}$. The blue curve is the QFT solution from Ref.~\cite{Kofman:2004yc}, and the red curve is the solution obtained from the analytical model, \eq{produc}, of the rate of energy loss to open strings.}
\label{lowv}
\end{figure}

As explained above, bremsstrahlung leads to rapid deceleration even at large distances, so that a D-brane pair undergoing scattering will generally have a nonrelativistic relative velocity by the time that the separation is small enough for open string pair production to be relevant.  For the remainder of this appendix we therefore focus on open string pair production in the case $\eta \ll 1$.  (The case $\eta \gg 1$ is treated in Ref.~\cite{McAllister:2004gd}.)
Using the asymptotic relation
\be
{\theta_{10}^4\left(0,is\right)\over \eta_{\te{D}}^{12}\left(is\right)}=  16 + {\cal{O}}(e^{-2 \pi s})
\ee for $s\gg1$, the production rate $\Gamma_{\text{prod}}(\tau)$ may be written as
\bea
\Gamma_{\text{prod}}(\tau)
\approx \zeta \xi^2(\tau){{16 V_{\te{D}p}} \over {(2 \pi)^p}}  \left(v \over {2\pi } \right)^{(p+1)/2} \exp\Bigl(-{{b^{2}+v^{2}\tau^{2}} \over{2 v}}\Bigr)&\text{for $\eta\ll1$}\,.\label{rate}
\eea
Knowing the production rate of open strings,  it is straightforward to obtain the rate at which open string production absorbs energy.
The open string number density and energy density are related by \cite{McAllister:2004gd}
\bea\label{rhoerel}
\rho_{\text{open}} \approx
 \frac{\sqrt{b^{2}+v^{2} \tau^{2}}}{2\pi\eta}\,\nu_{\text{open}}&\text{for $\sqrt{b^{2}+v^{2} \tau^{2}}\gg\eta$\,.}
\eea
Using \eq{rate} and  \eq{rhoerel} we find that the rate of change of the energy in open strings is
\bea\label{produc}
\dot{\rho}_{\text{prod}}\approx  \frac{\zeta \xi^2(\tau)\pi^4}{(2\pi)^{(3p+3)/2}}{\sqrt{b^{2}+\tau^{2}}}\,\eta^{(p-11)/2}\exp\left(\eta-{b^{2}+\tau^{2}\over2\eta}\right)~&\text{for $\sqrt{b^{2}+v^{2} \tau^{2}}\gg\eta$\,.}
\label{produc_a}
\eea
Strong evidence for the energy loss rate given in \eq{rate} comes from comparing a simulation based on \eq{rate} to a simulation that directly incorporates the instantaneous production rate, as in Ref.~\cite{Kofman:2004yc}.  We display the resulting trajectories, which agree very well over many orbits, in Figure~\ref{lowv}.

The open string production rate given \eq{rate} is strictly applicable only for trajectories with constant rapidity: when $\eta$ varies with time, the T-dual configuration used in the derivation of the annulus amplitude no longer involves a constant angle.
However, Figure \ref{lowv} shows that interpreting \eq{rate} as a function of varying $\eta$ is an excellent approximation in the corresponding field theory model, and we expect it to be applicable in level-truncated string field theory as well.

In scattering processes that lead to production of a high density of excited open strings, the late-time dynamics is not amenable to analytic treatment:
the full problem involves quantum production, cascade decay, and
annihilation of a diverse population of open strings, as well as the growth of spatial inhomogeneities.
Moreover, the angular momentum stored in open strings can have a significant impact on the evolution.
A rough expectation is that the D-branes will initially enter an elliptical orbit, but the orbital timescale is often much longer than the Jeans time, so that the homogeneous approximation fails before many orbits are completed.
This is a system similar to Branonium, as described in Ref.~\cite{Burgess:2003qv}, with the addition of a high density of strings stretched between the branes.

Thus far we have assumed that open string pair production can be described as a spatially homogeneous process, but a finite density of pairs necessarily involves inhomogeneities on short lengthscales.
Consider the nucleation of a single open string pair at a given location on an initially homogeneous D$p$-brane that is moving relativistically toward another D$p$-brane, which we take to be fixed.  The creation of the open string pair absorbs momentum from a small patch of the moving D$p$-brane, so that the  D$p$-brane cannot remain spatially homogeneous.  The resulting acceleration triggers bremsstrahlung, inducing further deceleration, so that a roughly spherical shock-wave of acceleration and bremsstrahlung propagates outward  (within the $p$ spatial dimensions of the brane) from the nucleation point.  Recalling that the critical acceleration given in \eq{aH} is $a_{\te{H}}= ({2\sqrt{2}\pi \gamma^{3}})^{-1}$, which is very small in string units for an ultrarelativistic D-brane, we find that the creation of an open string pair  leads to an expanding shock that may be strong enough to trigger emission of massive closed strings.
A proper treatment of inhomogeneous pair creation would be interesting, but is beyond the scope of this work.

\section{Critical Frequency for D-dimensional Synchrotron Radiation} \label{wkb}

In this appendix we give a few details of the calculation of the critical frequency $\omega_{\te{c}}$  for
 scalar and vector\footnote{An explicit calculation of the tensor spectrum of an ultrarelativistic particle in a spacetime of arbitrary dimension $D$ is beyond the scope of this work.} synchrotron radiation in a spacetime of even dimension $D$.
We will show that for $\gamma \gg 1$ the critical frequency $\omega_{\te{c}}$ does not depend on the spacetime dimension $D$.
Our analysis closely follows Ref.~\cite{Breuer}, where the special case $D=4$ was considered.

The wave equation for a massless scalar $\phi$ with coupling $\alpha_{D}$ to the stress-energy  of a source
takes the form
\be
\Box \phi(x_{\mu} )
= \alpha_{D}T\,.
\ee
The field $\phi$ can be written as (see Ref.~\cite{ Avery:1989})
\be
\phi(t, r, \theta_{i})=\sum_{\ell_{1}=0}^{\infty}\sum_{\ell_{2}=0}^{\ell_{1}}...\sum_{\ell_{D-3}=0}^{\ell_{D-4}}\sum_{m=-\ell_{D-3}}^{\ell_{D-3}}\int d\omega ~ \phi_{\ell_{1}...\ell_{D-3}m}(r) H_{\ell_{1}...\ell_{D-3}m}(\theta_{1},...,\theta_{D-2})e^{-i\omega t}\,,
\ee
where $\theta_{i}$ are the $D-2$ angular directions \{$\theta_{i}\in[0,2\pi)~\forall~i<D-2,~\theta_{D-2}\in[0,\pi)$\}, and $H_{\ell_{1}...\ell_{D-3}m}$ are  hyperspherical harmonics on $S^{D-2}$.
The $D$-dimensional d'Alembertian is
\be
\Box= -{\partial^2\over\partial t^2}+r^{2-D} {\partial\over\partial r}\Bigl(r^{D-2}{\partial\over\partial r}\Bigr)+{1\over r^{2}}\Delta_{\te{S}^{D-2}}\,,
\ee
where the hyperspherical harmonics satisfy $\Delta_{S^{D-2}}H_{\ell_{1}...\ell_{D-3}m}=-\ell_{1}(\ell_{1}+D-3)H_{\ell_{1}...\ell_{D-3}m}$.
Using the substitution $\phi_{\ell_{1}...\ell_{D-3}m}=u_{\ell_{1}...\ell_{D-3}m}/r^{(D-2)/2}$, the equation of motion for the  Fourier components takes the form
\be
\Bigl(\partial_{r}^{2}+\omega^{2}-U(r)\Bigr)u_{\ell_{1}...\ell_{D-3}m}=0\,,
\ee
for $r>r_{0}$, where $\omega^{2}=m^{2}\omega_0^{2} = m^2 v^2/r_{0}^2$, and
\be
U(r)={\ell_{1}(\ell_{1}+D-3)\over r^{2}}+{(D-4)(D-2)\over 4 r^{2}}\,.
\ee
Our goal is to
obtain the critical frequency $\omega_c$ above which the spectrum falls off exponentially, and we therefore take $m \gg 1$.
A solution at $r_{0}$ can be obtained by the WKB method (for $\omega\gg\omega_{0}$):
\be
u_{\ell_{1}...\ell_{D-3}m}(r_{0})={1\over\sqrt{U(r_0)-\omega^{2}}}\exp\bl-\int_{r_{0}}^{r_{+}}dr~\sqrt{U(r)-\omega^{2}}\br\, ,
\ee
where $r_{+}$ is the classical turning point: $U(r_{+})=\omega^{2}$. The radiated power is related to  $u(r)$ by $P\propto |u(r_{0})|^{2}$, so that the leading exponential dependence of the power spectrum is given by
\be \label{leadingexp}
{dP\over d\omega}\propto \exp\bl-2\int_{r_{0}}^{r_{+}}dr~\sqrt{U(r)-\omega^{2}}\br\, .
\ee
It remains to evaluate the integral in order to find $\omega_{\te{c}}$. We can solve $U(r_+)=\omega^2$ for $r_{+}$:
\be
r_{+} \approx \frac{r_{0}}{v} \sqrt{\frac{2\ell_{1}+D/2-2}{m}-1} +\ldots\,
\ee
(note that $\ell_{1} \ge m$ and $m \gg 1$, so $r_+$ is real).
Using this result in
\be
\int_{r_{0}}^{r_{+}}dr~\sqrt{V(r)-\omega^{2}} \approx {1\over 2}(r_{+}-r_{0})\,\sqrt{V(r_{0})-\omega^{2}}\,,
\ee
we conclude that
for $\ell_1 \gg 1$ and $m\gg 1$, the leading contribution to the critical frequency $\omega_c$ is {\it{independent}} of the dimension $D$.
Thus, in the relativistic limit the critical frequency for radiation by a point particle is given by
\be \label{Omegac}
\omega_{\te{c}}=\omega_{0}\gamma^{3}+\mathcal{O}\Bigl((D-3) \omega_0 \gamma^2\Bigr)\,,
\ee and the leading term takes the same form as in four-dimensional electromagnetism.

\end{document}